\documentclass[twocolumn]{aastex631}
\usepackage[T1]{fontenc}
\usepackage{amsmath}
\usepackage{amsthm}
\usepackage{amssymb}
\usepackage{graphicx}
\usepackage{physics}
\usepackage{color}
\usepackage{hyperref}

\shorttitle{HERA--Roman Cross-Correlations}
\shortauthors{La Plante, Mirocha, et al.}

\received{2022 May 25}

\revised{2022 Dec 21}

\accepted{2022 Dec 22}

\published{2023 Feb 13}

\submitjournal{ApJ}

\newcommand\zreion{\texttt{zreion}}
\newcommand\mAB{m_{\rm{AB}}}
\newcommand\xLAE{f_{\rm{LAE}}}
\newcommand\fLAE{f_{\rm{LAE}}}
\newcommand\Lya{\rm{Ly}\alpha}

\newcommand\MUV{M_\mathrm{UV}}
\newcommand\AUV{A_\mathrm{UV}}
\newcommand\mab{m_\mathrm{AB}}
\newcommand\Ha{\rm{H}\alpha}
\newcommand{\OIII}{[\rm{O} \ \textsc{iii}]}
\newcommand\Mstell{M_\star}

\begin{document}

\title{Prospects for 21\,cm--Galaxy Cross-Correlations with HERA and the Roman High-Latitude Survey}

\author[0000-0002-4693-0102]{Paul La Plante}
\affiliation{Department of Astronomy, University of California, Berkeley, CA 94720, USA}
\affiliation{Berkeley Center for Cosmological Physics, University of California, Berkeley, CA 94720, USA}
\affiliation{Department of Computer Science, University of Nevada, Las Vegas, NV 89154, USA}
\affiliation{Nevada Center for Astrophysics, University of Nevada, Las Vegas, NV 89154, USA}

\author[0000-0002-8802-5581]{Jordan Mirocha}
\affiliation{Department of Physics and McGill Space Science Institute, McGill University, Montreal, Canada}

\author[0000-0002-1712-737X]{Adélie Gorce}
\affiliation{Department of Physics and McGill Space Science Institute, McGill University, Montreal, Canada}

\author[0000-0002-3950-9598]{Adam Lidz}
\affiliation{Center for Particle Cosmology, Department of Physics and Astronomy, University of Pennsylvania, Philadelphia, PA 19104, USA}

\author[0000-0002-5400-8097]{Aaron Parsons}
\affiliation{Department of Astronomy, University of California, Berkeley, CA 94720, USA}

\correspondingauthor{Paul La Plante}
\email{paul.laplante@unlv.edu}

\begin{abstract}
  The cross-correlation between the 21\, cm field and the galaxy distribution is a potential probe of the Epoch of Reionization (EoR). The 21\,cm signal traces neutral gas in the intergalactic medium and, on large spatial scales, this should be anti-correlated with the high-redshift galaxy distribution which partly sources and tracks the ionized gas.  In the near future, interferometers such as the Hydrogen Epoch of Reionization Array (HERA) are projected to provide extremely sensitive measurements of the 21\,cm power spectrum. At the same time, the Nancy Grace Roman Space Telescope (Roman) will produce the most extensive catalog to date of bright galaxies from the EoR. Using semi-numeric simulations of reionization, we explore the prospects for measuring the cross-power spectrum between the 21\,cm and galaxy fields during the EoR. We forecast a 12$\sigma$ detection between HERA and Roman, assuming an overlapping survey area of 500 deg$^2$, redshift uncertainties of $\sigma_z = 0.01$ (as expected for the high-latitude spectroscopic survey of Ly$\alpha$-emitting galaxies), and an effective Ly$\alpha$ emitter duty cycle of $f_\mathrm{LAE} = 0.1$. Thus the HERA--Roman cross-power spectrum may be used to help verify 21\,cm detections from HERA. We find that the shot-noise in the galaxy distribution is a limiting factor for detection, and so supplemental observations using Roman should prioritize deeper observations, rather than covering a wider field of view. We have made a public GitHub repository containing key parts of the calculation, which accompanies this paper: \url{https://github.com/plaplant/21cm_gal_cross_correlation}.
\end{abstract}

\section{Introduction}
\label{sec:intro}

The Epoch of Reionization (EoR) is one of the last remaining frontiers in observational cosmology. The EoR is the time period when the first luminous sources in the universe formed and gradually photo-ionized neutral hydrogen in the surrounding intergalactic medium (IGM). This is thought to occur roughly 0.5--1 Gyr after the Big Bang. This large-scale transition of the universe has yet to be fully observed: in particular, the redshift evolution of the average ionization fraction, $x_i(z)$, and the overall topology of the reionization process remain highly uncertain. Measurements of the ionization history and its spatial fluctuations would help pin-down the properties of early star-forming galaxies, as well as cosmological parameters such as the optical depth of cosmic microwave background (CMB) photons to electron scattering, $\tau$. As such, providing measurements of the EoR has important ramifications for multiple fields in astrophysics and cosmology.

Observationally, radio interferometers are attempting to measure the 21\,cm signal of neutral hydrogen \citep{madau_etal1997}. The hyperfine transition of neutral hydrogen atoms emits and absorbs radiation with a rest wavelength of 21\,cm, which can imprint an excess or deficit in brightness relative to the CMB backlight. 
During the EoR, the 21\,cm signal is expected to have significant fluctuations due to the presence of large ionized regions, because once ionized the hydrogen in the IGM no longer emits a 21\,cm signal. While many experiments seek the 21\,cm power spectrum (e.g., the Low Frequency Array, LOFAR, \citealt{lofar}, the Murchison Widefield Array, MWA, \citealt{mwa}, and the Owens Valley Long Wavelength Array, \citealt{ovlwa}), in this work we focus on the Hydrogen Epoch of Reionization Array (HERA, \citealt{deboer_etal2017}), currently under construction in the Karoo desert of South Africa. Once completed, HERA will be the most sensitive instrument to date measuring the 21\,cm signal, and is expected to provide the first statistically significant detection of the 21\,cm power spectrum \citep{deboer_etal2017}.

At the same time, wide-field infrared telescopes are planning to observe a large ensemble of high-redshift galaxies directly, e.g., Euclid\footnote{\url{https://www.cosmos.esa.int/web/euclid/euclid-survey}} and the Nancy Grace Roman Space Telescope \citep[Roman,][]{wfirst}. In this work, we focus in particular on Roman, expected to launch in 2026, as its wide-field surveys are deeper than the nominal Euclid survey. Among other scientific goals, Roman is slated to observe 2200 deg$^2$ of the sky with both imaging and spectroscopy, the so-called High Latitude Survey (HLS), with imaging and spectroscopic campaigns referred to as the HLIS and HLSS, respectively. Although this survey is designed to detect galaxies with redshift $1 \lesssim z \lesssim 3$, it can also detect galaxies at redshifts of $z \gtrsim 5$ via the Lyman break technique \citep{Steidel1996}. These high-redshift galaxies include some of the sources thought to be responsible for ionizing the universe during the EoR. In principle, the galaxy field measured in such a wide-field survey and the 21\,cm signal should be statistically anti-correlated on large spatial scales during the EoR, at least in the expected case that large-scale overdensities are ionized first, i.e. in an ``inside-out'' reionization scenario. This anti-correlation is due to the galaxy field tracing highly biased regions of cosmic structure, whereas the 21\,cm signal comes predominantly from lower-density, neutral regions. As such, the 21\,cm and galaxy fields should be anti-correlated, at least on spatial scales large relative to the typical size of ionized bubbles.

The cross-power spectrum of the 21\,cm and galaxy fields is also interesting in principle as a means to validate measurements made of the 21\,cm auto power spectrum from HERA. Detecting the cross-power spectrum between the 21\,cm and an independent tracer will provide important evidence that should confirm inferences made from HERA measurements alone, such as information about size distribution of ionized bubbles and their evolution with redshift. Furthermore, at low redshift the only 21\,cm fluctuation measurements made to date have been in cross-correlation with galaxy and quasar catalogs, rather than as a 21\,cm auto power spectrum. For example, the recent results from the Canadian Hydrogen Intensity Mapping Experiment (CHIME\footnote{\url{https://chime-experiment.ca/en}}) report a detection of the cross-power spectrum between the 21\,cm signal and galaxies and quasars from eBOSS \citep{CHIME2022}. Additionally, H~\textsc{i} intensity maps from $z \sim 1$ made by the Green Bank Telescope have been used in cross-correlation with galaxy surveys to measure the hydrogen abundance and bias parameters \citep{chang_etal2010,masui_etal2013,switzer_etal2013,wolz_etal2022}. As such, modeling and measuring the 21\,cm--galaxy cross-power spectrum at high redshift is useful independent of the 21\,cm auto power spectrum.


Previous studies \citep{furlanetto_lidz2007,Wyithe2007,lidz_etal2009,vrbanec_etal2020} have looked at similar prospects for cross-correlation between 21\,cm experiments and wide-field galaxy surveys. We move past this previous work by making cross-spectrum forecasts for the upcoming HERA and Roman data for the first time. Important, we also explicitly account for the ``foreground wedge'' \citep{datta_etal2010,pober_etal2013,parsons_etal2014}. That is, previous studies accounted only for foregrounds preventing measurements of Fourier modes with long-wavelength line-of-sight components, i.e., low-$k_\parallel$ modes. In fact, the frequency dependence of the instrumental response of an interferometer leads to mode-mixing and this corrupts some high $k_\parallel$ modes as well. Nevertheless, this contamination is expected mostly to occupy a wedge-shaped region in the $k_\parallel$-$k_\perp$ plane. This further degrades the prospects for cross-power spectrum measurements, as discussed in this work.
Furthermore, we include a detailed treatment of the Roman observations. Specifically, we account for
the nominal magnitude and flux limits for the HLIS and HLSS, a complete handling of redshift uncertainties, and projections for the joint overlap area between HERA and Roman, now that such information is available \citep{wfirst,Dore2018,Wang2022}. Finally, we model how the cross-correlation signal varies as a function of ionization history and galaxy properties. Although these conclusions may be partly tied to the particular seminumeric simulation of reionization employed, these model variations nevertheless have important implications for the detectability and interpretation of the cross-spectrum signal. 


We organize the rest of this paper as follows. In Section~\ref{sec:methods}, we described the semi-numeric simulations used for this study. In Section~\ref{sec:results}, we present the primary results of our work. In Section~\ref{sec:detectability}, we discuss the feasibility of detection for upcoming 21\,cm surveys. In Section~\ref{sec:discussion}, we explore ways that observations can improve on the fiducial measurement strategies. Finally, in Section~\ref{sec:conclusion}, we provide a summary and avenues for future research. Throughout this work, we assume a $\Lambda$CDM cosmology with parameters consistent with the Planck 2018 results \citep{planck2018}.

\section{Methods}
\label{sec:methods}

In this section, we describe the simulations used to explore the EoR and to model the cross-correlation power spectrum between the 21\,cm and galaxy fields. We expect the 21\,cm and galaxy fields to be well-correlated on relatively large scales ($\gtrsim 1$ $h^{-1}$Mpc) based on previous work \citep{lidz_etal2009}. Thus we opt to simulate a large volume with moderate resolution, as this helps capture features of the field on the scales probed by upcoming observations. We begin by describing the semi-numeric reionization method that we use, followed by our galaxy modeling. We also include a discussion of the various observational systematics present for both the 21\,cm measurements as well as the galaxy field.


\subsection{21\,cm Modeling}
\label{sec:zreion}
Accurately simulating the EoR is a computationally difficult problem. The complex interplay between dark matter, baryons, and photons necessitates employing $N$-body methods, hydrodynamics, and radiative transfer to capture the variety of physical effects self consistently. Furthermore, the formation of luminous objects on sub-kpc scales has implications for the ionization state of the IGM on scales of tens of Mpc, which requires a tremendous amount of dynamic range in the simulation. For now, even state-of-the-art simulation packages are incapable of handling such a tremendous workload. Furthermore, for the present study, we are most interested in the behavior of the IGM on relatively large scales, and so many of the small-scale details inside of individual galaxies are not relevant. Thus, we opt to use a semi-numeric scheme for simulating reionization, which captures the main features in a reliable fashion while remaining relatively cheap computationally to allow for an exploration of the parameter space of various ionization histories. Specifically, we make use of the \zreion\ semi-numeric reionization code \citep{battaglia_etal2013a}, which has previously been applied to modeling the 21\,cm signal in \citet{laplante_etal2014}, \citet{laplante_ntampaka2019}, and \citet{laplante_etal2020}.

The central \textit{Ansatz} of \zreion\ is that the matter density field $\delta_m(\vb{r})$ and the redshift at which a particular portion of the IGM is reionized $z_\mathrm{re}(\vb{r})$ are correlated on large scales. We begin by defining the matter overdensity field $\delta_m(\vb{r})$ as:
\begin{equation}
\delta_m(\vb{r}) \equiv \frac{\rho_m(\vb{r}) - \bar{\rho}_m}{\bar{\rho}_m},
\end{equation}
where $\bar{\rho}_m$ is the mean matter density, and the equivalent ``overdensity'' field for the redshift of reionization $\delta_z(\vb{r})$ is
\begin{equation}
\delta_z(\vb{r}) \equiv \frac{\qty[z_\mathrm{re}(\vb{r}) + 1] - \qty[\bar{z} + 1]}{\bar{z} + 1},
\label{eqn:deltaz}
\end{equation}
where $\bar{z}$ is the mean value for the $z_\mathrm{re}(\vb{r})$ field. The \zreion\ method expresses the correlation between $\delta_m(\vb{r})$ and $\delta_z(\vb{r})$ as a scale-dependent bias factor $b_{zm}(k)$, which is defined as:
\begin{equation}
b_{zm}^2(k) \equiv \frac{\ev{\delta_z^*\delta_z}_k}{\ev{\delta_m^*\delta_m}_k} = \frac{P_{zz}(k)}{P_{mm}(k)}. \label{eq:bias}
\end{equation}
We parametrize this bias factor using two parameters $k_0$ and $\alpha$, and express the bias as a function of Fourier wavenumber $k$:
\begin{equation}
b_{zm}(k) = \frac{b_0}{\qty(1 + \frac{k}{k_0})^\alpha}.
\label{eqn:bias}
\end{equation}
We use the value of $b_0 = 1/\delta_c = 0.593$, where $\delta_c$ is the critical overdensity in spherical collapse halo models. Given a cosmological density field, \zreion\ relies only on the value of the parameters $\{\bar{z},k_0,\alpha\}$ to determine the redshift of reionization $z_\mathrm{re}(\vb{x})$. The midpoint of reionization---defined as the time at which the universe is 50\% ionized by volume---is largely determined by the mean value $\bar{z}$ (though is not identically equal to the parameter), and the duration is controlled by $k_0$ and $\alpha$.

For the current work, we run a series of dark-matter only simulations which
contain 1024$^3$ particles in a cubic volume with length
$L = 2~h^{-1}\mathrm{Gpc}$ on a side, which corresponds to an angular extent of $\theta \approx 18$ deg at $z = 8$. This is about twice the extent of the instantaneous field-of-view (FoV) of HERA \citep{deboer_etal2017}. We first generate a set of initial
conditions based off of transfer functions obtained from CAMB
\citep{lewis_etal2000}. Given these initial conditions, we use second-order
Lagrangian perturbation theory (2LPT) to evolve the particle positions as a
function of redshift. Although 2LPT does not capture the non-linear evolution of
particles to the same extent as $N$-body methods, the results are sufficiently
accurate on the scales of interest \citep{scoccimarro1998}. To generate the
ionization field, we use 2LPT to evolve the particles to the midpoint of
reionization $\bar{z}$. To compute the matter density field
$\delta_m(\vb{x})$, we use triangular shaped clouds (TSC) to deposit the particles in
the simulation onto a regular cubic lattice. With the density field in
hand, we apply a fast Fourier transform (FFT) and apply the bias from
Equation~(\ref{eqn:bias}) to $\delta_m(\vb{k})$ to compute
$\delta_z(\vb{k})$. After applying an inverse FFT, we use
Equation~(\ref{eqn:deltaz}) to compute $z_\mathrm{re}(\vb{x})$. To convert from
this quantity at a particular redshift $z_0$ into an ionization field
$x_i(\vb{x},z_0)$, we set all values of the ionization field to have a value of
1 where $z_\mathrm{re}(\vb{x})$ is greater than $z_0$ (meaning that portion of
the volume was reionized at an earlier time), and 0 otherwise.

Once we have the ionization and density fields, we can convert to the 21\,cm brightness temperature by using \citep{madau_etal1997}:
\begin{equation}
\delta T_b(\vb{r}, z) = T_0(z) \qty[1 + \delta_m(\vb{r})]\qty[1 - x_i(\vb{r})],
\label{eqn:tb}
\end{equation}
where the quantity $T_0(z)$ is
\begin{align}
T_0(z) &= 26\qty(\frac{T_S - T_\gamma}{T_S})\qty(\frac{\Omega_b h^2}{0.022}) \notag \\
&\qquad \times \qty[\qty(\frac{0.143}{\Omega_m h^2})\qty(\frac{1 + z}{10})]^{\frac{1}{2}}.
\label{eqn:t0}
\end{align}
In this equation, $T_S$ is the spin temperature of neutral hydrogen and $T_\gamma$ is the temperature of the CMB. It is common in the literature to assume the spin temperature to be coupled to the  temperature of the gas $T_\mathrm{gas}$. This is thought to occur via X-ray photo-heating, perhaps driven by X-ray binaries in early galaxies, such that the gas becomes substantially hotter than $T_\gamma$ before it is significantly ionized \citep{furlanetto_etal2006}. For contrast, it is also useful to consider a ``cold reionization'' scenario in which there is no such heating. In this case, we suppose that $T_\mathrm{gas}$ cools adiabatically between when the gas thermally decouples from the CMB (near $z \sim 200$) and reionization.  This makes $T_\mathrm{gas}$ smaller than $T_\gamma$, so the brightness temperature in Equation~(\ref{eqn:tb}) is negative and has a larger amplitude than in our fiducial scenario. We consider this case in Section~\ref{sec:s/n}.

\subsubsection{Changing the Ionization History}
\label{sec:ion_hist}
To explore the extent to which these results depend on the precise ionization history, we vary the parameters of our semi-numeric reionization model described in Section~\ref{sec:zreion} to produce alternate plausible histories. In addition to our ``fiducial'' scenario, we run a series of simulations that have a shorter reionization history with a comparable midpoint (the ``short'' scenario). We also run a simulation which has a slightly later midpoint of reionization (the ``late'' scenario),\footnote{The \texttt{zreion} parameters for the fiducial scenario are: $\bar{z} = 8$, $\alpha = 0.2$, and $k_0 = 0.9$ $h$Mpc$^{-1}$. The short scenario parameters are: $\bar{z} = 8$, $\alpha = 0.564$, $k_0 = 0.185$ $h$Mpc$^{-1}$. The late scenario parameters the same as the fiducial scenario, but with $\bar{z} = 7$.} which can improve our understanding of how the results are impacted by the midpoint of reionization occurring at redshift values that are not covered by the Roman grism (as mentioned above in Section~\ref{sec:galaxy_surveys}).

\begin{figure}[t]
  \centering
  \includegraphics[width=0.48\textwidth]{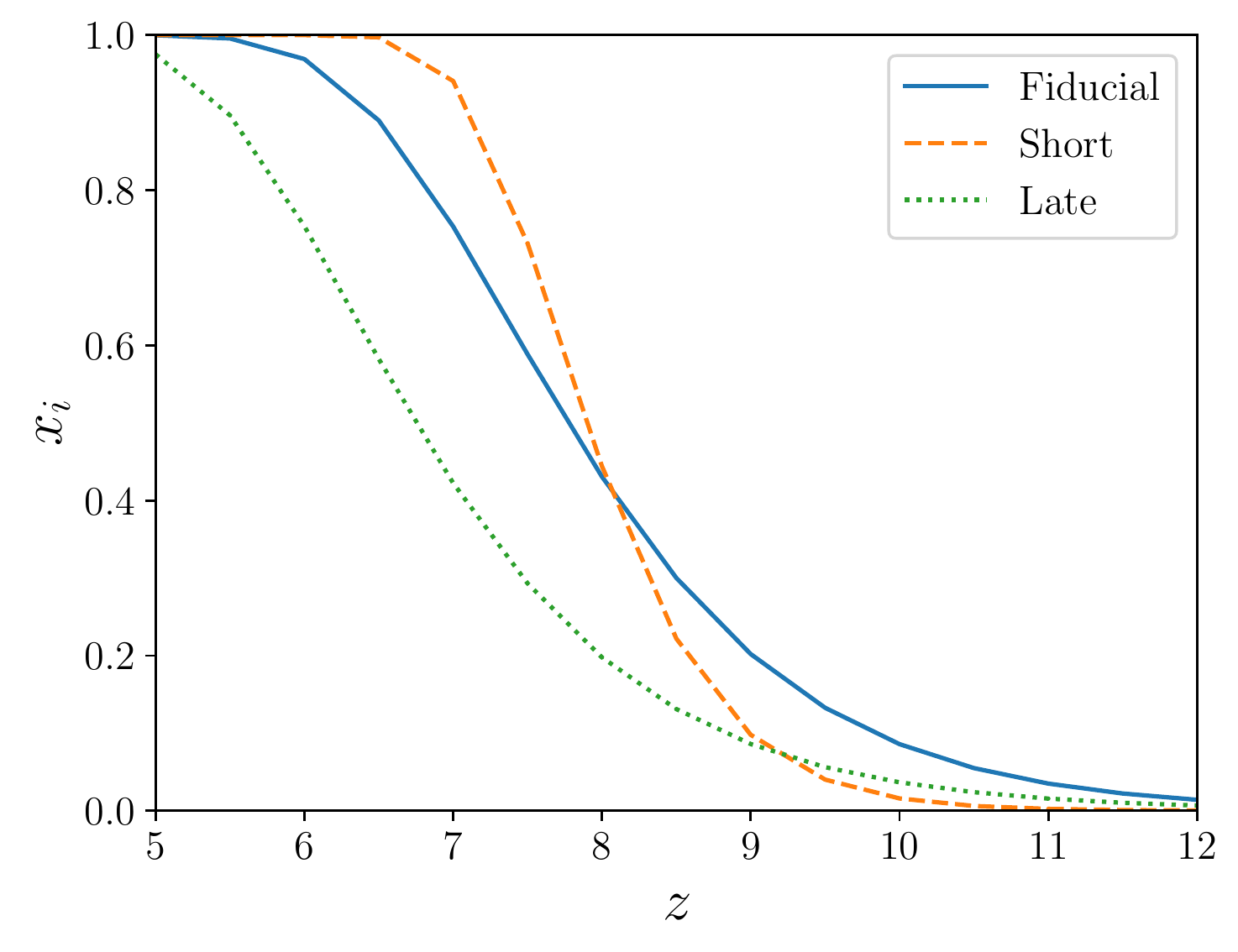}
  \caption{\textbf{The global ionization fraction $x_i$ as a function of redshift $z$.} We show this quantity for the three scenarios explored in this work. As mentioned in Section~\ref{sec:ion_hist}, these different histories help show the extent to which our forecasts are affected by details specific to one particular history.}
  \label{fig:xhist}
\end{figure}

Figure~\ref{fig:xhist} shows the ionization fraction as a function of redshift for the different ionization histories described. In all cases, we use the same values of the galaxy bias parameters described below in Section~\ref{sec:galaxy_models}. As discussed more there, we do not include any explicit connection between the ionization field and the galaxy field besides the fact that both are treated as biased tracers of the matter density field. However, given that we are primarily interested in relatively bright and biased galaxies in the HLS observations on large scales, we do not expect this simplification to be a significant source of error. 

In the results below in Sections~\ref{sec:results} and \ref{sec:detectability}, we present results for all three of the ionization histories. In general, the amplitude of the cross-spectrum increases as the duration of reionization decreases. This is due to a larger amplitude of the 21\,cm fluctuations on large scales for these shorter scenarios, which is a general feature of the semi-numeric model used (and explored more in \citealt{laplante_etal2014}).

\subsection{Galaxy Modeling} \label{sec:galaxy_models}
Like reionization, galaxy formation is a rich and complicated physical process
that involves the interactions between dark matter, baryons, and radiation over
many decades in length scale. As a result, self-consistent simulations of galaxy formation and reionization -- especially in the large volumes relevant to cross-correlations -- remain exceedingly challenging \citep{iliev_etal2006,trac_cen2007,oshea_etal2015,ocvirk_etal2016,lewis_etal2022}. Given that we are most interested in the
large-scale correlation between the 21\,cm and galaxy fields, we use a linear bias model to rapidly generate a galaxy abundance field rather than simulating it from first principles, though the linear bias values we adopt are themselves based on semi-analytic and hydrodynamical models.

We first describe our model for star formation in high-$z$ galaxies in Section~\ref{sec:galaxy_basics} and the resulting predictions for the galaxy bias and abundance. Then, we describe our approach to $\Lya$ emission in Section~\ref{sec:galaxy_lya}, with an emphasis on the likelihood that galaxies detected in the Roman high-latitude \textit{imaging} survey (HLIS) are also detected in the high-latitude \textit{spectroscopic} survey (HLSS).

\subsubsection{Galaxy Properties} \label{sec:galaxy_basics}
The bias of galaxies -- one of two key inputs to our model -- can be readily computed from simulations using an expression analogous to Eq. \ref{eq:bias}, but using the galaxy overdensity field in place of the reionization redshift field. This is precisely what has been done for the \textsc{BlueTides} simulations \citep{feng_etal2014,waters_etal2016}. In this work, we will employ both the \textsc{BlueTides} predictions for the galaxy bias, as well as an efficient semi-empirical model implemented in \textsc{ares}\footnote{\url{https://github.com/mirochaj/ares}}. Both models have been designed to match the high-redshift rest-UV luminosity functions, and so are in good agreement with current datasets by construction. \textsc{ares} allows us to explore potential extensions to the HLS, for which galaxy bias predictions from simulations are not readily available. We summarize the most pertinent aspects of this model here briefly, and refer the interested reader to \citet{Mirocha2017} and \citet{Mirocha2020} for more details.

Our basic approach is similar to many semi-empirical models put forth in recent years. We assume that the star formation rate (SFR) in galaxies is driven by halo growth, $\dot{M}_{\ast} = f_{\ast}(M_h) \dot{M}_h$, where $f_{\ast}$ is a halo mass-dependent star formation efficiency. We derive halo mass accretion rates (MAR) under the assumption that halos evolve at fixed number density \citep[see][]{Furlanetto2017}, which provides a good match to MARs derived from numerical simulations \citep{Mirocha2021}. We assume $f_{\ast}$ is a double power-law, and calibrate its free parameters by jointly fitting to high-$z$ UV luminosity functions (UVLFs) and UV colours ($\beta$) from \citet{Bouwens2015} and \citet{Bouwens2014}. Two key assumptions remain. First, we adopt the \textsc{BPASS} version 1 single-star models \citep{Eldridge2009} with a stellar metallicity of $Z=0.004$\footnote{Note that our results are insensitive to this choice. Because the model is calibrated empirically, any change in $Z$ must be met with a commensurate change in $f_{\ast}$, which keeps the rest-ultraviolet emission roughly constant. See, e.g., Section~3.4 in \citet{Mirocha2017} for more discussion of this.}. Second, whereas many models in the literature neglect dust or adopt empirical relationships between dust attenuation and UV colour, we self-consistently forward model the full rest-ultraviolet spectrum of each model galaxy, with additional free parameters governing the dust production efficiency and scale length allowed to vary as well \citep[see][]{Mirocha2020}. This results in slightly different relationships between, e.g., $\MUV$ and the UV extinction $\AUV$ than are predicted from the relationship between infrared excess and $\beta$ at lower redshifts, such as that of \citet{Meurer1999}.

\begin{figure*}[t]
  \centering
  \includegraphics[width=0.98\textwidth]{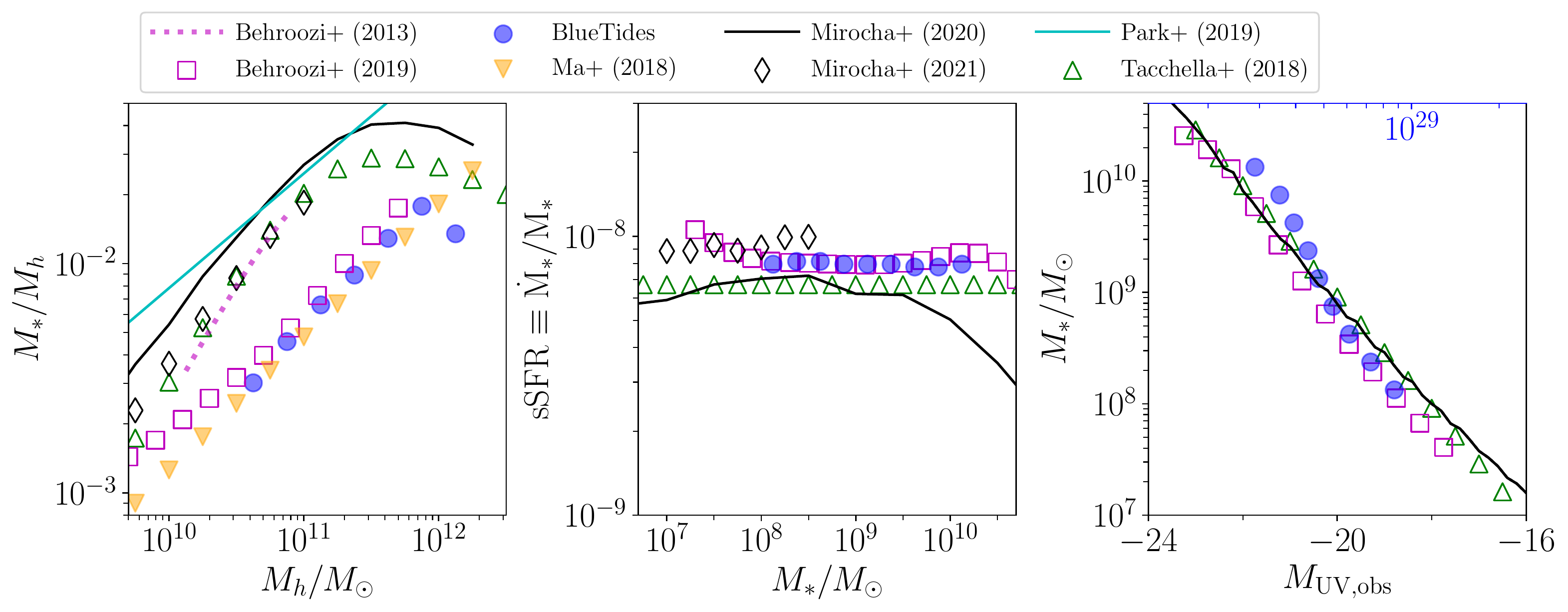}
  \caption{\textbf{Predictions for galaxy properties at $z\sim 8$ from several models.} \textit{Left:} Stellar mass -- halo mass relation, \textit{Center:} Specific star formation rate (sSFR) vs. stellar mass. \textit{Right:} Relationship between stellar mass and observed UV magnitude (upper axis is $L_{1600}/[\rm{erg}\ \rm{s}^{-1} \ \rm{Hz}^{-1}]$ for \textsc{BlueTides}). Note that not all datasets are shown in every panel.}
  \label{fig:galaxy_properties}
\end{figure*}

For semi-empirical models like this, for which there is no simulation box, one must derive the galaxy bias as a weighted integral over the halo mass function, $dn/dm$, and halo bias, $b_h$,
\begin{equation}
    b_g(\geq m_{\min};z) = \frac{\int_{m_{\min}}^\infty b_h(m;z) \dv{n}{m} \dd{m}}{\int_{m_{\min}}^\infty \dv{n}{m} \dd{m}}
\end{equation}
where the minimum mass is related to the limiting magnitude of the survey, $m_{\min} = m_{\min}(m_{\rm{AB,lim}})$. We use a \citet{tinker_etal2008} mass function throughout, and adopt their fitting formula for the halo bias as well \citep{Tinker2010}.

In Figure~\ref{fig:galaxy_properties}, we compare the basic properties of galaxies in the \textsc{ares} semi-empirical model (solid black) to many others from the literature\footnote{Note that we show two versions of the \textsc{ares} model: a default approach using the mean halo MAR computed following \citet{Furlanetto2017}, and another in which halos histories are extracted from $N$-body simulations \citep{Mirocha2021}. The latter agrees more with \citet{Tacchella2018}, who also anchor their semi-empirical model to $N$-body simulations. The reduced SMHM is due to the steeper growth histories of simulated halos, thus making them brighter and bluer, allowing one to reduce the overall normalization of the SFE \citep{Mirocha2021}.}. The most striking differences among this set of models occur in the stellar mass--halo mass (SMHM) relation (left), while the specific star formation rate (sSFR) of galaxies (center) and stellar mass--UV magnitude relation (right) are much more similar. At a glance, it is the semi-empirical models that generally predict higher stellar masses at fixed halo mass \citep{Behroozi2013a,Behroozi2013b,Tacchella2018,Park2019,Mirocha2020}, as well as more not shown here \citep[e.g.][]{Sun2016}, while simulation-based estimates like \textsc{bluetides} \citep{feng_etal2015,feng_etal2016} and, e.g., FIRE-2 \citep{Ma2018}, exhibit lower $M_{\ast}/M_h$ ratios at fixed $M_h$, more in line with \textsc{UniverseMachine} predictions \citep{Behroozi2019} and other similar semi-empirical models built on $N$-body simulations \citep[e.g.][]{Moster2018}. However, one can also find examples in the literature of very detailed \textit{ab initio} simulations which predict higher SMHM ratios than \textsc{BlueTides} or \textsc{FIRE} \citep[e.g., \textit{Renaissance}, \textsc{SERRA};][]{Xu2016,Pallottini2022}, so the source of the differences here remain (as far as we know) unclear.


\begin{figure}[t]
  \centering
  \includegraphics[width=0.49\textwidth]{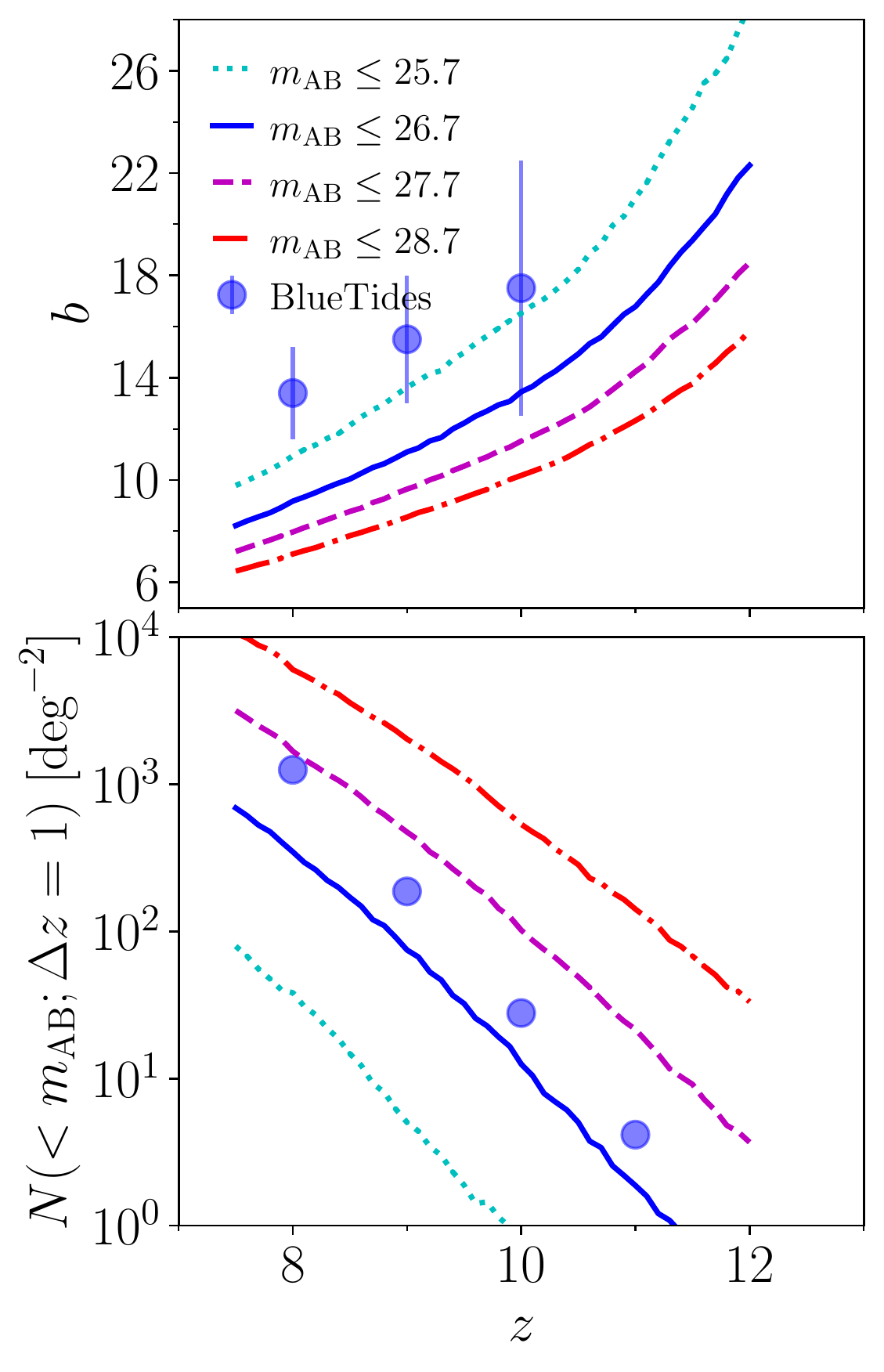}
  \caption{\textbf{Predictions for the galaxy bias and abundance at high redshifts.} \textit{Top:} Galaxy bias as a function of apparent UV magnitude cut, $\mAB$, computed with \textsc{ares} (lines) compared to \textsc{bluetides} (points), which adopted the nominal HLS value of $\mAB=26.7$. \textit{Bottom:} Galaxy surface densities over the same redshift interval.}
  \label{fig:bias_models}
\end{figure}

To our knowledge, only \citet{waters_etal2016} explicitly provide predictions for the bias of Roman HLS sources. From Figure~\ref{fig:galaxy_properties}, we expect the bias of sources in \textsc{BlueTides} to be higher than the \textsc{ares} models described earlier in this section: because there is good agreement between predictions for the specific star formation rate (sSFR) as a function of stellar mass among all models (middle panel), and reasonably good agreement also in predictions for the $\MUV$--$\Mstell$ relation (right), we conclude that at fixed stellar mass, the models predict very similar intrinsic \textit{and} dust-reddened luminosities. As a result, the SMHM relation will dominate differences in galaxy bias predictions. Indeed, this is the case, as we show in Figure~\ref{fig:bias_models} (top), along with predictions for the abundance of high-$z$ galaxies (bottom). Here, we show the \textsc{BlueTides} predictions as well as \textsc{ares} models with four different magnitude cuts: the nominal HLS limiting magnitude of 26.7 (solid blue), as well as scenarios a magnitude shallower (dotted cyan) and one and two magnitudes deeper (dashed and dash-dotted curves, respectively). As expected from Figure~\ref{fig:galaxy_properties}, these models differ non-trivially in their predictions: while \textsc{BlueTides} predicts $b(z=8) \simeq 13.5$, the \citet{Mirocha2020} models predict $b(z=8) \simeq 9$, with more rapid evolution at high redshifts than a linear extrapolation of \textsc{BlueTides} would suggest. In the bottom panel, we compare predictions for the surface density of galaxies as a function of redshift. One again there are noticeable differences, at the $\sim 2-3$x level.

The differences between model predictions for the SMHM relation (and thus galaxy bias) are certainly interesting and warrant attention \citep[see, e.g., Section~4.1 in][for some more discussion]{Tacchella2018}. In this work, however, we will remain agnostic about which of these models (if any) are correct, and instead use their differences to motivate a plausible range of possibilities to explore in our cross-correlation forecast. Because \textsc{BlueTides} has provided predictions for the bias of HLS sources explicitly, we will explore this as one possible scenario, and use the \textsc{ares} models as a contrasting case, for which we can efficiently generate alternative scenarios for different survey parameters or galaxy properties. For additional comparisons of the \textsc{BlueTides} and \textsc{ares} predictions, see Appendix \ref{sec:appendix}.

Finally, once we have the linear bias in hand, we construct the galaxy field from the matter density field $\delta_m$ through the relation:
\begin{equation}
\delta_g(\vb{k},z) = b_g(z)\delta_m(\vb{k},z).
\label{eqn:deltag}
\end{equation}
Note that given the relatively large values for the bias $b_g$, as seen in Figure~\ref{fig:bias_models}, the simulated galaxy distribution does sometimes reach non-physical values in the simulation, where $\delta_g < -1$. Although this is a limitation of our current treatment, we do not expect these values to cause significant inaccuracies in the resulting predictions given that most of the cross-spectrum sensitivity comes from large spatial scales where a linear biasing model is a good approximation.

\subsubsection{Nebular Emission} \label{sec:galaxy_lya}
So far, we have only considered the bias and abundance of galaxies detected in at least one band in the high-latitude imaging survey. However, a key question is whether or not the galaxies detected in the imaging survey will also be detected spectroscopically, since redshift uncertainties $\sigma_z \gtrsim 0.1$ (e.g., photometric redshifts alone) are expected to prevent a cross-correlation detection \citep{furlanetto_lidz2007}.
Specifically, if the redshift uncertainties in the galaxy survey are too large, this prevents measuring higher $k_\parallel$ modes in the galaxy distribution, while 21\,cm surveys generally lose the low $k_\parallel$ modes to foreground contamination. For 21\,cm-galaxy cross-power spectrum measurements, it is thus crucial to obtain good spectroscopic redshifts in the galaxy survey, as this will help ensure that each survey measures some common Fourier modes.
We discuss the impact of measurement uncertainties in more detail in Section~\ref{sec:galaxy_surveys}, and here focus on whether galaxies bright enough to be detected in imaging ought to also be bright enough in their $\Lya$ emission to be detected in the spectroscopic survey.

We take a simple approach and assume photo-ionization equilibrium in the H\textsc{ii} regions of galaxies. This allows us to relate the luminosity of $\Lya$ to the recombination rate ($\sim 2/3$ of recombinations result in $\Lya$ photons), which is in turn related to the intrinsic ionizing photon production rate, one of the main predictions of the galaxy model. The implementation in \textsc{ares} is described in more detail in Section~2.2.2 of \citet{Sun2021}.

\begin{figure}[t]
  \centering
  \includegraphics[width=0.49\textwidth]{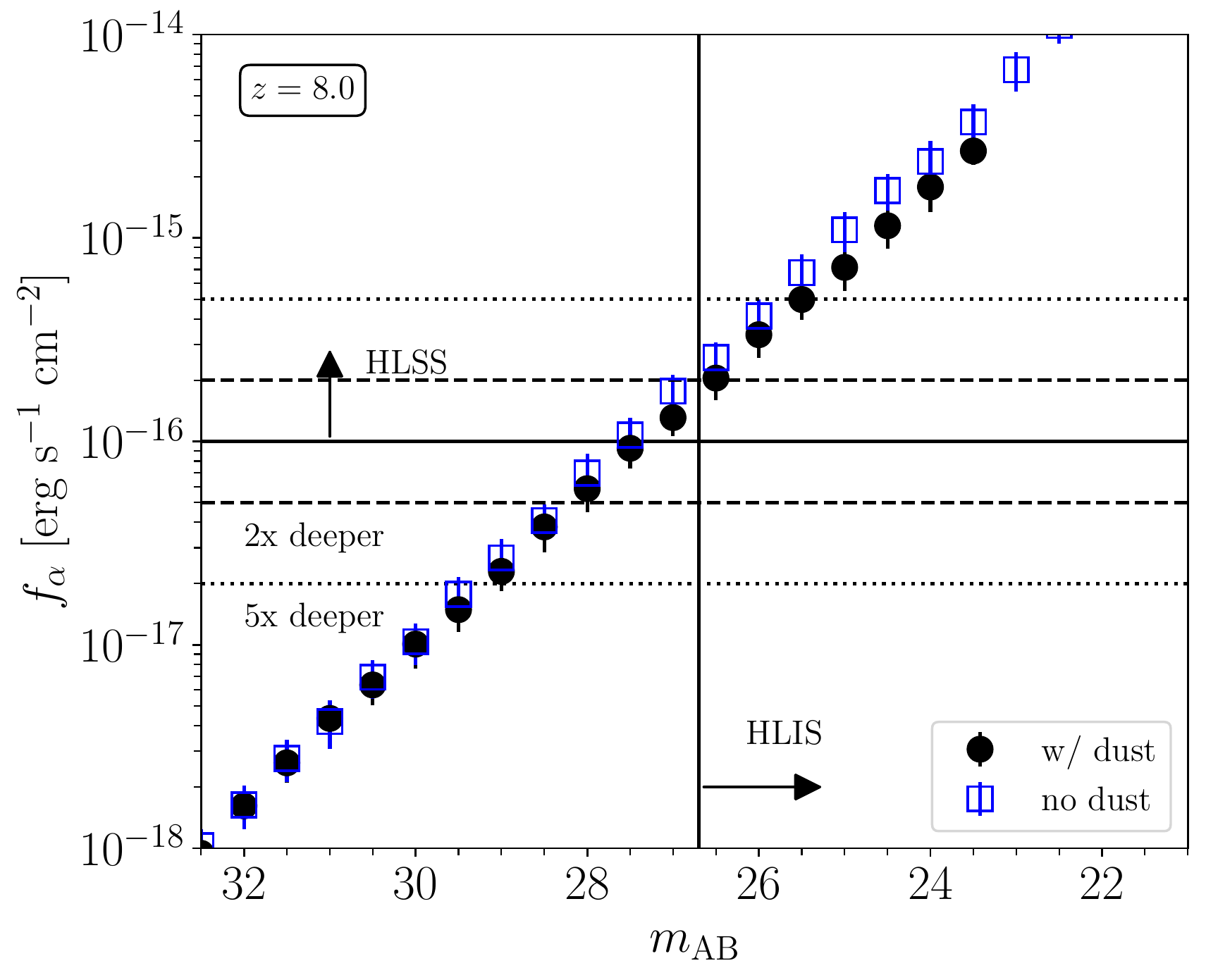}
  \caption{\textbf{Relationship between intrinsic Ly-$\alpha$ line luminosity and galaxy UV magnitude.} Solid lines indicate the nominal imaging (vertical) and spectroscopic (horizontal) survey depths. The dashed and dotted curves indicate factor of 2 and 5 deviations from the nominal spectroscopic survey, respectively. All HLIS sources detected in the rest-UV continuum should be detected in the HLSS with the nominal sensitivity, neglecting IGM transmission effects.}
  \label{fig:lya_mab}
\end{figure}

\begin{figure*}[t]
  \centering
  \includegraphics[width=0.95\textwidth]{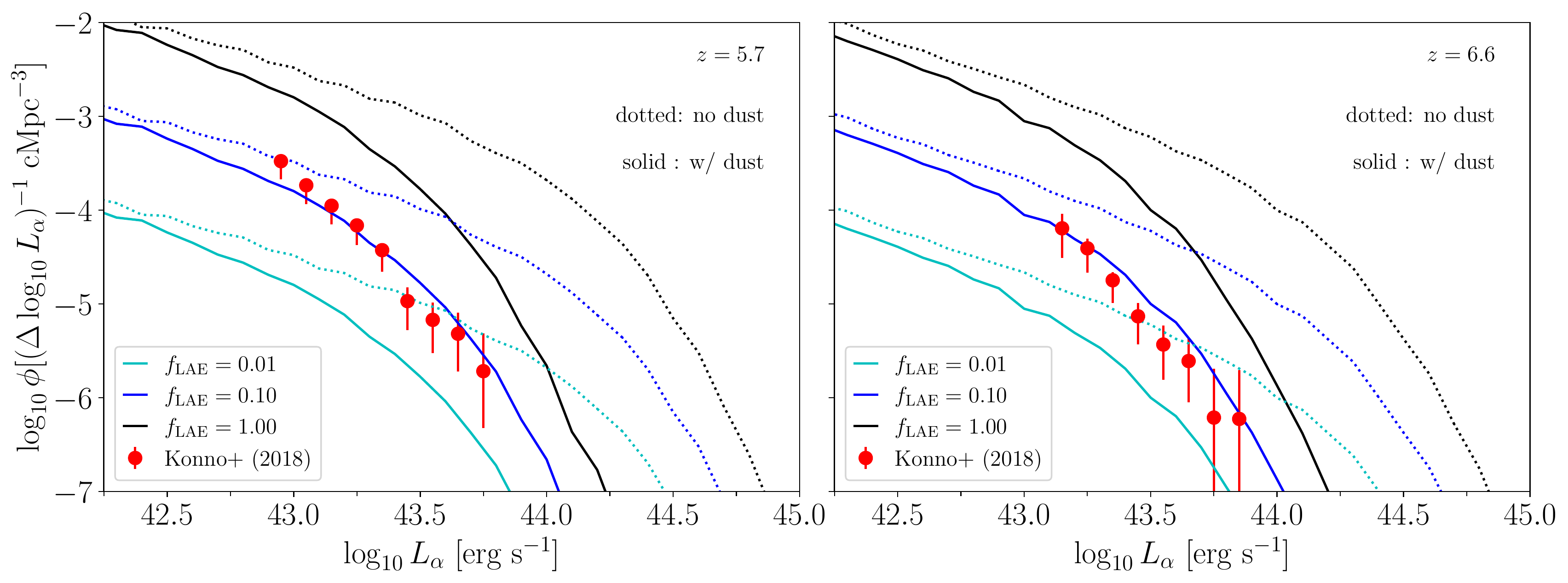}
  \caption{\textbf{The LAE galaxy luminosity function predicted by \textsc{ares} and measured from SILVERRUSH \citep{konno_etal2018}}. As can be seen, an effective LAE duty cycle of $f_\mathrm{LAE} = 0.1$ with dust reddening agrees reasonably well with the data at both $z \sim 5.7$ and $z \sim 6.6$. We use this value as our fiducial value for the rest of the analysis, but include predictions for $f_\mathrm{LAE} = 1$ and $f_\mathrm{LAE} = 0.01$ as additional points of comparison.}
  \label{fig:lyalf}
\end{figure*}

In Figure \ref{fig:lya_mab}, we show our predictions for the relationship between intrinsic $\Lya$ line luminosity and apparent UV magnitude at $z=8$, with nominal sensitivity limits for the HLIS and HLSS overlaid for reference (vertical and horizontal lines, respectively). For the flux limits, we adopt the nominal sensitivity quoted in \citet{Wang2022}, and include factor of 2x and 5x deeper/fainter surveys for reference as well. Objects detected in the upper right quadrant of this plot are detected in both imaging \textit{and} spectroscopy, while objects in the lower right or upper left quadrants are only detected in imaging \textit{or} spectroscopy, respectively. We can see clearly that an object detected in imaging should also be detectable spectroscopically via its $\Lya$ emission. This remains the case regardless of whether one includes dust or not (circles v. squares), since dust here attenuates the UV continuum and $\Lya$ similarly. Setting the dust contents to zero by hand will boost the overall number of galaxies, as well as the magnitude of the brightest galaxy, but not the relationship between $\Lya$ flux and $\mAB$. Additionally, each of the galaxies in the HLSS will be observed from multiple roll angles, and so to reduce the incidence of false positives, galaxies will be required to be observed from at least 3 different rolls.

Figure~\ref{fig:lya_mab} also tells us that galaxies that are slightly too faint to be detected in imaging may still be detectable in the HLSS. However, given that Roman uses a grism, the expectation is to only extract spectra for known sources identified in imaging.
The result shown in Figure~\ref{fig:lya_mab} is not entirely a surprise. The nominal HLS limiting magnitude and flux limits are chosen to optimize low-$z$ science, e.g., the ability to detect $1 \lesssim z \lesssim 3$ galaxies both in imaging of the rest-optical continuum (with some contamination from lines) and the rest-optical emission lines themselves, like $\Ha$ and $\OIII$. Because the continuum of star-forming galaxies is relatively flat from the ultraviolet through the optical, and recombination results in $\sim 1$ $\Lya$ photon for every $\Ha$ photon, it makes sense that sensitivities set for low-$z$ science goals are about right for detecting high-$z$ galaxies in $\Lya$ and UV continuum as well.

Before moving on, note that so far we have neglected IGM transmission effects, effectively assuming that every HLS galaxy resides in a very large fully ionized bubble. In all that follows, we adopt a LAE duty cycle or fraction, $\fLAE$, which reduces the number of spectroscopically detected galaxies but \textit{not} the luminosity of any individual object. In practice, the LAEs observable by Roman will have some spatial and luminosity dependence primarily owing to absorption from neutral regions in the IGM. However, for the sake of simplicity in this initial study, we instead use an overall factor and defer a more realistic treatment to future work.
\footnote{Because the optical depth to Ly$\alpha$ is so large close to the line center, even an LAE in an ionized bubble will generally be attenuated. For instance, if the intrinsic line were a simple Gaussian of some width, the blue half of the line will typically be lost, even in a large ionized bubble. In future work we plan to account for attenuation of LAEs due to the IGM explicitly, but for now use the overall factor $f_\mathrm{LAE}$ to include these various physical effects.}

Figure~\ref{fig:lyalf} shows a comparison between the LAE galaxy luminosity function predicted from \textsc{ares} and the measurements from the SILVERRUSH survey \citep{konno_etal2018}. We include our galaxy models both with and without dust reddening, and plot different factors of $f_\mathrm{LAE}$. As can be seen from the figure, a value of $f_\mathrm{LAE} = 0.1$ and including dust reddening agrees reasonably well with the available measurements at $z \sim 5.7$ and $z \sim 6.6$. Note that our overall factor $f_\mathrm{LAE}$ is not inferred the same way as $f_\mathrm{duty}$ which sometimes appears in the interpretation of these measurements. Constraints on $f_\mathrm{duty}$ are typically based on clustering measurements: one infers a large-scale bias, which can be related to a minimum halo mass and thus a prediction for the total number density of halos. The value for $f_\mathrm{duty}$ is then computed as the number of LAEs divided by the number of expected halos. In our case, the number of LAEs in any $L_\alpha$ bin is reduced both by $f_\mathrm{LAE}$ and dust reddening. As a result, we don't need extreme values of $f_\mathrm{LAE} = 0.01$ unless dust is completely negligible.

In the following analysis, we choose $f_\mathrm{LAE} = 0.1$ as our default value, but also include predictions for $f_\mathrm{LAE} = 1$ and $f_\mathrm{LAE} = 0.01$ for the sake of comparison and understanding how this values impacts our final results. For example, $f_\mathrm{LAE} = 0.01$ may be a more realistic value at higher redshift, when the IGM becomes more neutral and the opacity to Ly$\alpha$ photons increases due to there being smaller neutral regions (and hence less time for photons to redshift out of the Ly$\alpha$ transition window).

\begin{figure*}[t]
\centering
\includegraphics[width=0.48\textwidth]{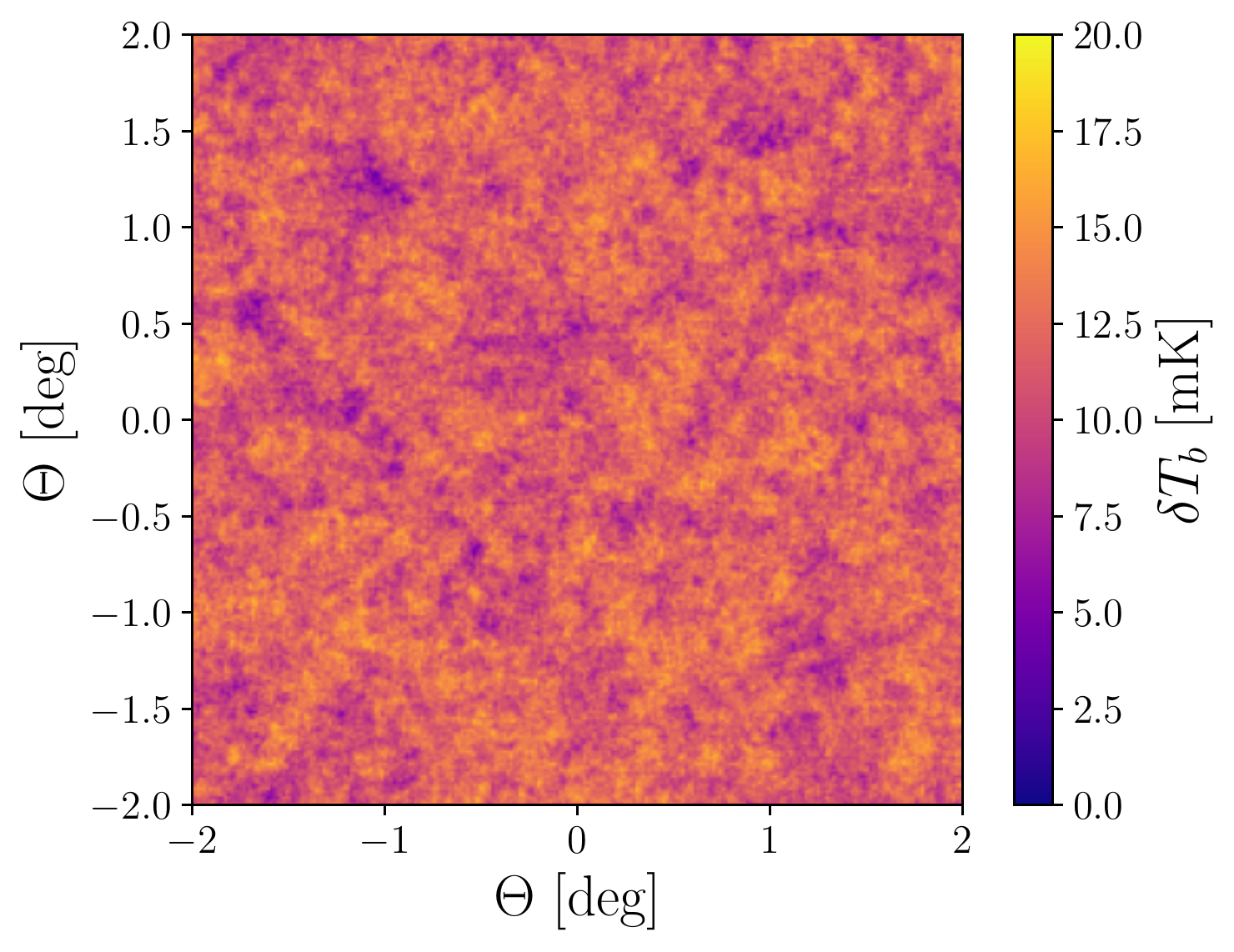}\hfill%
\includegraphics[width=0.48\textwidth]{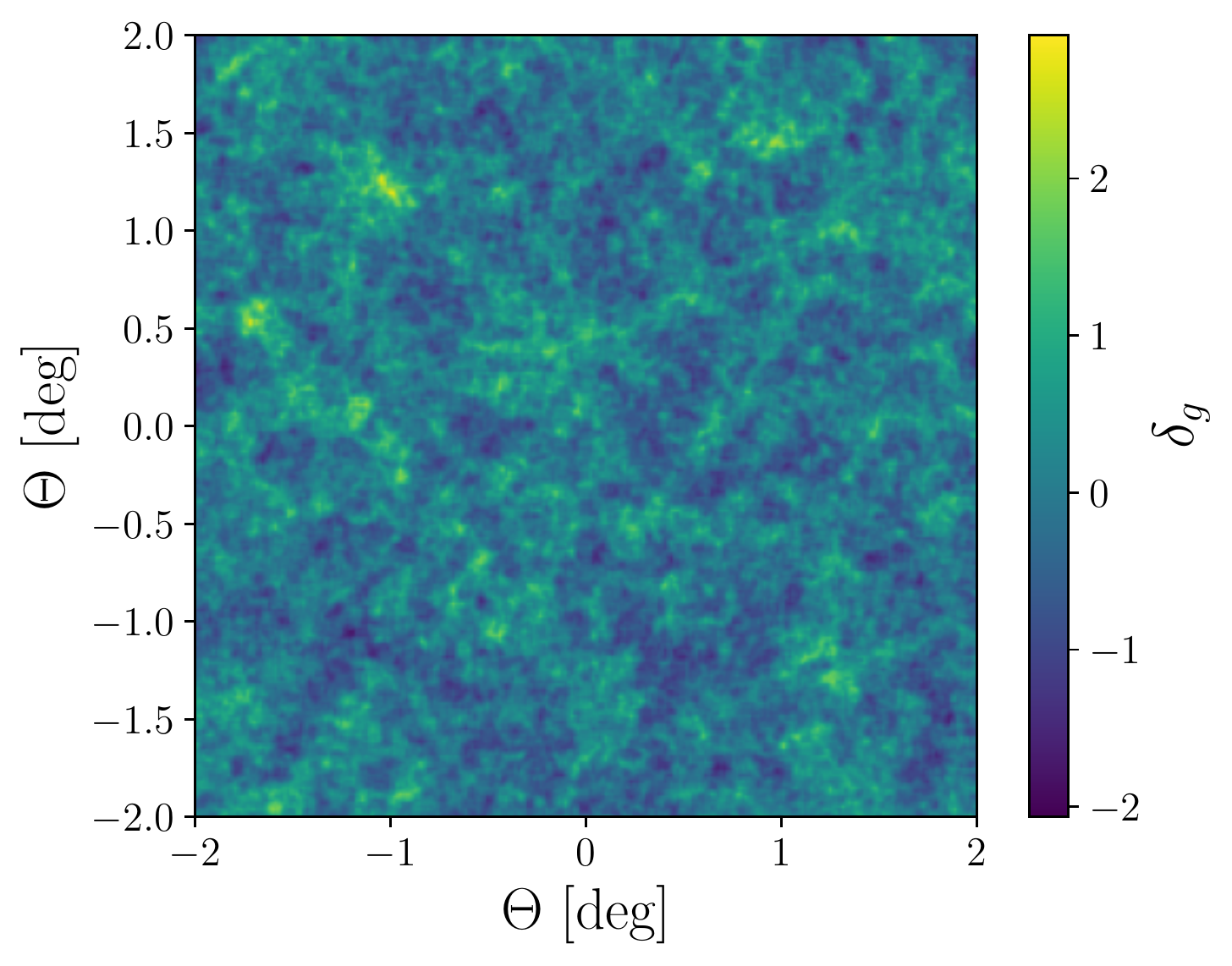}
\caption{\textbf{A visualization of our simulation volumes.} Left: the 21\,cm field generated from our simulations described in Section~\ref{sec:zreion}. This $2^\circ \times 2^\circ$ field represents about 1\% of the simulated sky area from one of our simulations. Right: a self-consistently generated galaxy field described in Section~\ref{sec:galaxy_models}. These two-dimensional slices are averaged over the redshift window $7.5 \leq z \leq 8.5$, which spans the midpoint of reionization for this realization. Although we show the two-dimensional field for illustration purposes, we compute the cross-spectrum in three dimensions to avoid catastrophic cancellation of Fourier modes contaminated by 21\,cm foregrounds.}
\label{fig:fields}
\end{figure*}

\subsection{The 21\,cm-galaxy cross-spectrum}
Once we have the 21\,cm brightness temperature field defined in Equation~(\ref{eqn:tb}) and the galaxy field defined in Equation~(\ref{eqn:deltag}), we are able to compute the cross-spectrum $P_\mathrm{21\times gal}$. For the sake of comparing the results with observations, we compute this quantity as a function of $k_\perp$ and $k_\parallel$, i.e., the Fourier modes which lie in the plane of the sky and along the line of sight, respectively. Given the relatively small angular extent of our simulations, we use the flat-sky approximation when computing power spectra. We average over all modes that contribute to a given cylindrical wavenumber bin, described by $(k_\perp, k_\parallel)$:
\begin{equation}
P_\mathrm{21\times gal}(k_\perp, k_\parallel) = \ev{\delta T_b^*\delta_g}_{k_\perp,k_\parallel}.
\label{eqn:xspec}
\end{equation}
We first construct a pair of lightcones---one each for the 21\,cm field and the galaxy field---in a self-consistent fashion using our 2LPT simulations described above. Once we have these lightcones, we apply FFTs  and then compute the cross-power spectrum as described above.

Figure~\ref{fig:fields} shows a visualzation of the 21\,cm field and the galaxy fields as generated be the semi-numeric models described above in Secs.~\ref{sec:zreion} and \ref{sec:galaxy_models}. These fields are generated from a realization of our fiducial ionization history by averaging over the respective lightcones between $7.5 \leq z \leq 8.5$. Note that in practice when computing the cross-spectrum defined above in Equation~(\ref{eqn:xspec}), we compute the FFT in three dimensions and do not average prior to computing this cross-spectrum. This approach ensures that we retain the full Fourier space information and can accurately model the effect of avoiding modes which are contaminated by the 21\,cm foregrounds described below in Section~\ref{sec:21cm_obs}.

\subsection{Observational Effects}
\label{sec:obs}

\subsubsection{21\,cm Observations}
\label{sec:21cm_obs}
In general, 21\,cm observations from interferometers are subject to foreground
contamination that are many orders of magnitude larger than the cosmological
signal from reionization \citep{morales_wyithe2010,pober_etal2013}. The dominant
foreground is due to galactic synchrotron radiation from the Milky Way,
which roughly follows a power law in frequency and hence is smooth in Fourier
space. Na{\"\i}vely, these bright foregrounds should be restricted to small
$k_\parallel$ modes in Fourier space. However, the chromatic response of the
interferometer causes this foreground signal to scatter into high $k_\parallel$
modes, creating a ``wedge'' in $(k_\perp,k_\parallel)$ Fourier space
\citep{parsons_etal2012}.

Mathematically, the slope of the wedge $m(z)$ is implicitly a function of
baseline length and cosmology, and can be expressed as
\citep{thyagarajan_etal2015}:
\begin{equation}
m(z) \equiv \frac{k_\parallel}{k_\perp} = \frac{\lambda(z) D_c(z) f_{21} H(z)}{c^2(1+z)^2},
\label{eqn:slope}
\end{equation}
where $\lambda(z)$ is the wavelength of the 21\,cm signal at a given redshift
$z$, $D_c(z)$ is the comoving distance to that redshift, $f_{21}$ is the
rest-frame frequency of the 21\,cm signal, and $H(z)$ is the Hubble
parameter. Note that this form of the foreground wedge in
Equation~(\ref{eqn:slope}) accounts for maximal data contamination: physically,
the bright foreground contamination extends down to the horizon of the
interferometer beam. Improved calibration of interferometers may make it
possible to work ``inside the wedge'' at points in Fourier space where the
contamination from foregrounds may be less severe, but for the purposes of this
work we assume this maximal amount of contamination. To include this effect, we
explicitly track which portions of Fourier space for each redshift are subject
to this foreground contamination. For the redshift values relevant to the EoR,
$m \sim 3$, which leads to a significant amount of Fourier space being subject
to this contamination.

In addition to the foreground contamination, we include the effects of thermal
noise present in measurements from HERA. For power spectrum measurements, the
primary analysis pipeline for HERA uses the so-called ``delay transform'' of
radio interferometer visibilities to estimate the power spectrum. Under the
flat-sky approximation, the thermal noise contribution to the variance of the
power spectrum can be written as a function of the observational frequency $\nu$ and wavenumber $u$ as \citep{parsons_etal2014}:
\begin{equation}
P_{21}^\mathrm{noise}(u, \nu) \approx \frac{T_\mathrm{sys}^2 \Omega_p^2(\nu) X^2(\nu) Y(\nu)}{\Omega_{pp}(\nu) t_\mathrm{int} N_\mathrm{pol} N_\mathrm{bl}(u)},
\label{eqn:21cm_noise}
\end{equation}
where $T_\mathrm{sys}$ is the system temperature of the interferometer,
$\Omega_p = \int \dd[2]{\vb{l}} A(\vb{l})$ is the integral of the primary beam
of the antenna $A(\vb{l})$, and $\Omega_{pp} = \int \dd[2]{\vb{l}} A^2(\vb{l})$
is the integral of the square of the primary beam. $X(\nu)$ and $Y(\nu)$ are
factors that convert the ``observed units'' of the interferometer into
cosmological units \citep{furlanetto_etal2006}. $X(\nu)$ accounts for the
conversion in the plane of the sky:
\begin{equation}
X(\nu) \equiv \dv{r_\perp}{l} = D_M(\nu),
\end{equation}
where $D_M(\nu)$ is the transverse comoving distance ($D_M = D_c$ for a flat
universe); and $Y(\nu)$ accounts for the conversion along the line of sight:
\begin{equation}
Y(\nu) \equiv \dv{r_\parallel}{\nu} = \frac{c(1+z)}{H(z)\nu}.
\end{equation}
Note that $X(\nu)$ has units of comoving Mpc per radian, and $Y(\nu)$ has units
of comoving Mpc per Hz. Finally, $t_\mathrm{int}$ is the integration time of the
measurement, $N_\mathrm{pol}$ is the number of instrumental polarizations
used to estimate the power spectrum, and $N_\mathrm{bl}(u)$ is the number of baseline pairs at a given $u$, where $u$ denotes the physical separation of HERA dishes in units of observed wavelength. Each baseline of length $u$ samples wavenumbers with transverse components, $k_\perp$, according to $u(\nu) = D_M(\nu) k_\perp / 2\pi$. For HERA, we assume a system temperature of $T_\mathrm{sys} = 400$~K, an observation time of $t_\mathrm{int} = 200$ hours, and $N_\mathrm{pol} = 2$ for two independent linear polarizations. We assume an overall observing season of 1000 hr, which is distributed among 5 non-overlapping fields. Such an approach was taken in the recent re-analysis of Phase I HERA data \citep{h1c_idr3}. We discuss combining observations from different regions of the sky more below in Sec.~\ref{sec:cum_snr}.

\begin{figure}[t]
  \centering
  \includegraphics[width=0.48\textwidth]{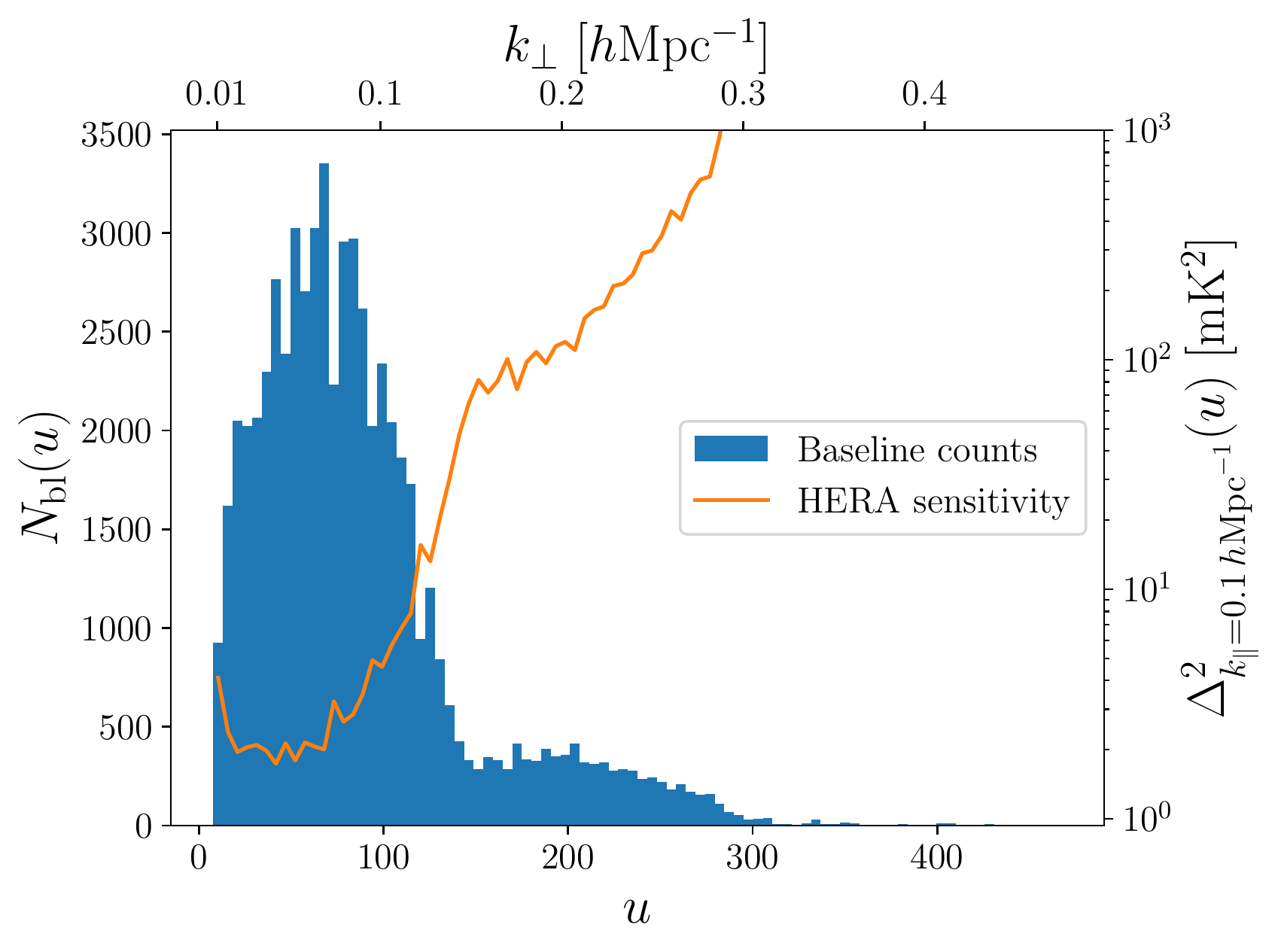}
  \caption{\textbf{The baseline distribution and sensitivity of HERA-350.} We plot the distribution of baselines for HERA as a function of wavenumber $u$ for the fully constructed 350-element array at $z=8$. The noise power spectrum in Equation~(\ref{eqn:21cm_noise}) is inversely proportional to the number of baselines which observe a particular $u$ mode on the sky, and is shown in orange. We plot the dimensionless noise power spectrum $\Delta^2(k) = k^3 P(k)/2\pi^2$ with a fixed value of $k_\parallel = 0.1$ $h$Mpc$^{-1}$, for 200 hr of observation (i.e., a single field).}
  \label{fig:nbl_dist}
\end{figure}

The number of baselines that observe a wavenumber $u$ depends on the configuration of the array. We show the baseline distribution for HERA in Figure~\ref{fig:nbl_dist}. HERA features many short baselines, so most of the baselines probe modes for which $u \lesssim 300$. We also show the noise floor for an observation, given by Equation~(\ref{eqn:21cm_noise}). For the current work, we limit ourselves to considering only the projected sensitivity for HERA. Next-generation radio telescopes, such as the SKA, are projected to provide data with even greater sensitivity. However, as we will see below in Section~\ref{sec:s/n}, the 21\,cm signal variance is dominated by cosmic variance of the modes used to make the measurements, rather than the instrumental uncertainty. As such, the additional sensitivity from the SKA may not significantly improve the measured significance of the 21\,cm auto-power spectrum. Nevertheless, the SKA may offer a significant improvement over HERA by being able to use more of the observable Fourier space, parametrized by the foreground wedge $m$ in Equation~(\ref{eqn:slope}). The SKA may also be able to observe more sky area that overlaps with the Roman HLS, yielding more joint sky coverage. It is also worth noting that given the improved sensitivity offered by the SKA, it may be possible to do a ``stacking'' type analysis in real (map) space, rather than Fourier space as assumed in this manuscript. It may be worthwhile to revisit some of these types of forecasts in the future to investigate the prospects for the SKA.

\subsubsection{Galaxy Surveys}
\label{sec:galaxy_surveys}
\begin{figure*}[t]
  \centering
  \includegraphics[width=0.98\textwidth]{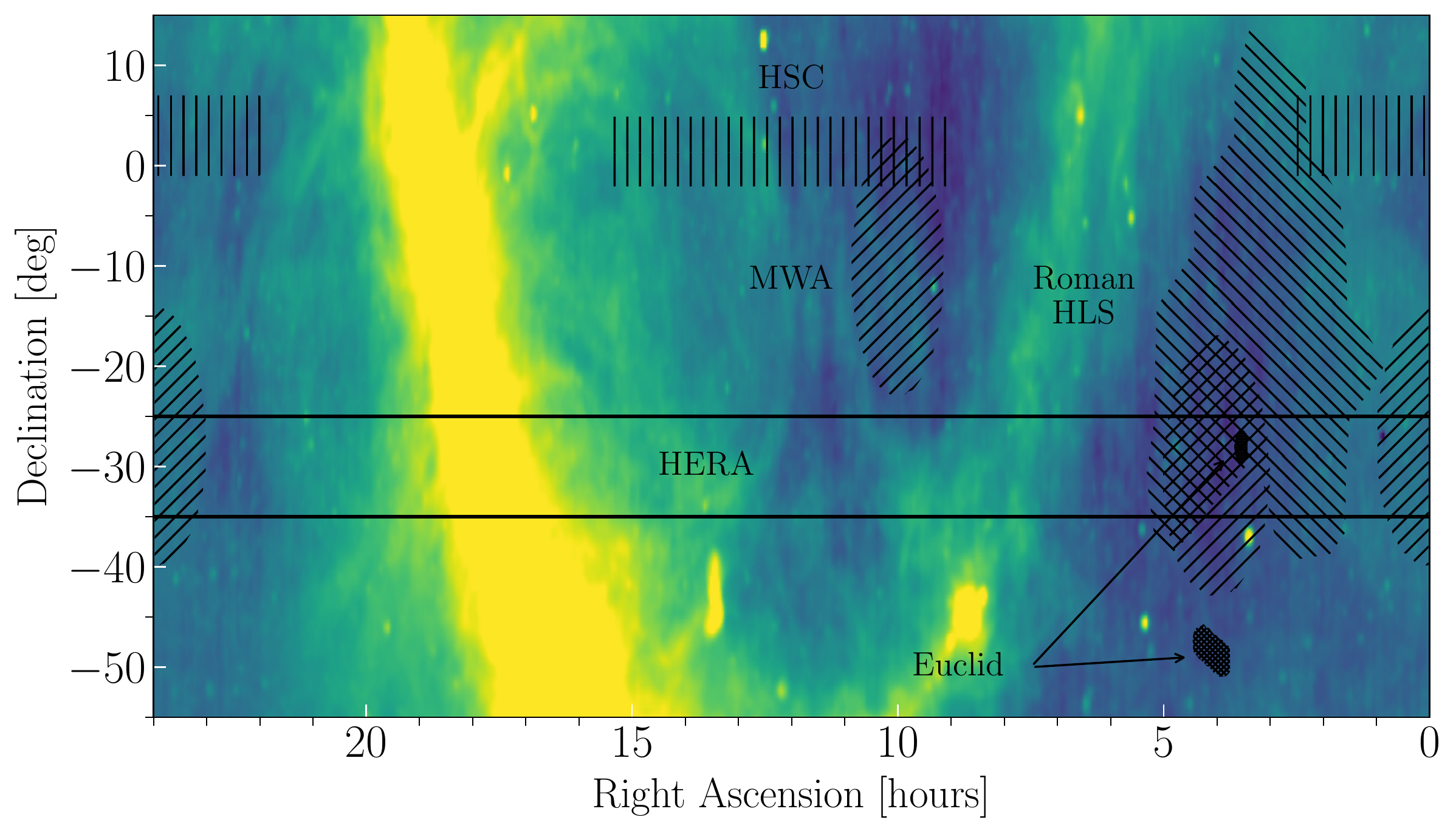}
  \caption{\textbf{Overlap of HERA and upcoming wide-field galaxy surveys.} HERA observes a stripe of constant declination, $\delta = 30^{\circ} \pm 5^{\circ}$, which partially overlaps with the nominal Roman HLS footprint (cross-hatching between $1 \lesssim \rm{RA}/\rm{hr} \lesssim 5$). Two Euclid deep fields are planned in the South that have sensitivity comparable to the HLS -- the Deep Field South and Deep Field Fornax -- but only the latter is in the HERA stripe. One of the three MWA EoR fields overlaps with Roman as well (circular cross-hatching), and the most northerly MWA field overlaps with LAE campaigns being conducted with Subaru HSC \citep[vertical hatching, see, e.g.,][]{Ouchi2018,Trott2021}. Background color-scale is the global sky model at 150 MHz \citep{GSM}.}
  \label{fig:survey_footprints}
\end{figure*}

Upcoming galaxy surveys will provide us the deepest and most comprehensive
catalog of high-redshift objects to date. Nevertheless, these objects are still
relatively faint and rare, and as such are only sparsely sampled. Furthermore,
the redshifts of the galaxies in these samples are only approximately known.
Assuming Poisson statistics and accounting for redshift uncertainties, the noise power spectrum for the galaxy field is:
\begin{equation}
P_\mathrm{gal}^\mathrm{noise}(k_\parallel) = \frac{e^{k_\parallel^2 \sigma_\chi^2}}{n_\mathrm{gal}},
\label{eqn:gal_noise}
\end{equation}
where $\sigma_\chi = c\sigma_z/H(z)$ is the co-moving distance uncertainty along the line of sight,
given a 1-$\sigma$ redshift precision of $\sigma_z$, and $n_\mathrm{gal}$
is the comoving number density of galaxies. The HLS in Roman is expected to have
a total survey area of 2200 deg$^2$ \citep{wfirst}. However, the HLS footprint
is not expected to fully overlap with the regions of the sky surveyed by
HERA. For the purposes of this analysis, we assume an overlapping sky region of 500 deg$^2$, as is expected given the nominal HLS footprint (see Figure~\ref{fig:survey_footprints}). Note that the planned Euclid deep fields are comparable to the HLS in depth. Unfortunately, only one lies in the HERA stripe, and is relatively narrow in the sky area it covers. We discuss trade-offs between depth and area further in Section~\ref{sec:discussion}.

Also vital to the success of any cross-correlation effort are precise redshift measurements. Photometric redshift estimates are expected to yield uncertainties of $\sigma_z \sim 0.5$ using the Lyman-break technique \citep{Bouwens2015,finkelstein2016}, while we expect uncertainties $\sigma_z \gtrsim 0.1$ to preclude a cross-correlation detection \citep{lidz_etal2009}. Given the high redshifts of interest, a successful HERA--Roman cross-correlation requires spectroscopically determined redshifts. Fortunately, the reference spectroscopic survey \citep{Wang2022} expects $\sigma_z \simeq 0.001 (1 + z)$, and so redshift measurements in principle should not be a limiting factor. We explore the effect of redshift uncertainty explicitly in Section~\ref{sec:s/n} below.

Given the sparsity of strong rest-frame ultraviolet lines in star-forming galaxies, $\Lya$ is likely the only line to be detected in Roman spectroscopy at the redshifts of interest. Furthermore, grism spectroscopy will only be extracted at the locations of sources detected in imaging. As a result, we have two requirements: (i) that galaxies be detected via the Lyman break technique, and (ii) that $\Lya$ is brighter than the sensitivity of the spectroscopic survey. Though there are potential systematic observing issues, we defer a discussion of these to Section~\ref{sec:galaxy_systematics}.

\begin{figure}[t]
  \centering
  \includegraphics[width=0.48\textwidth]{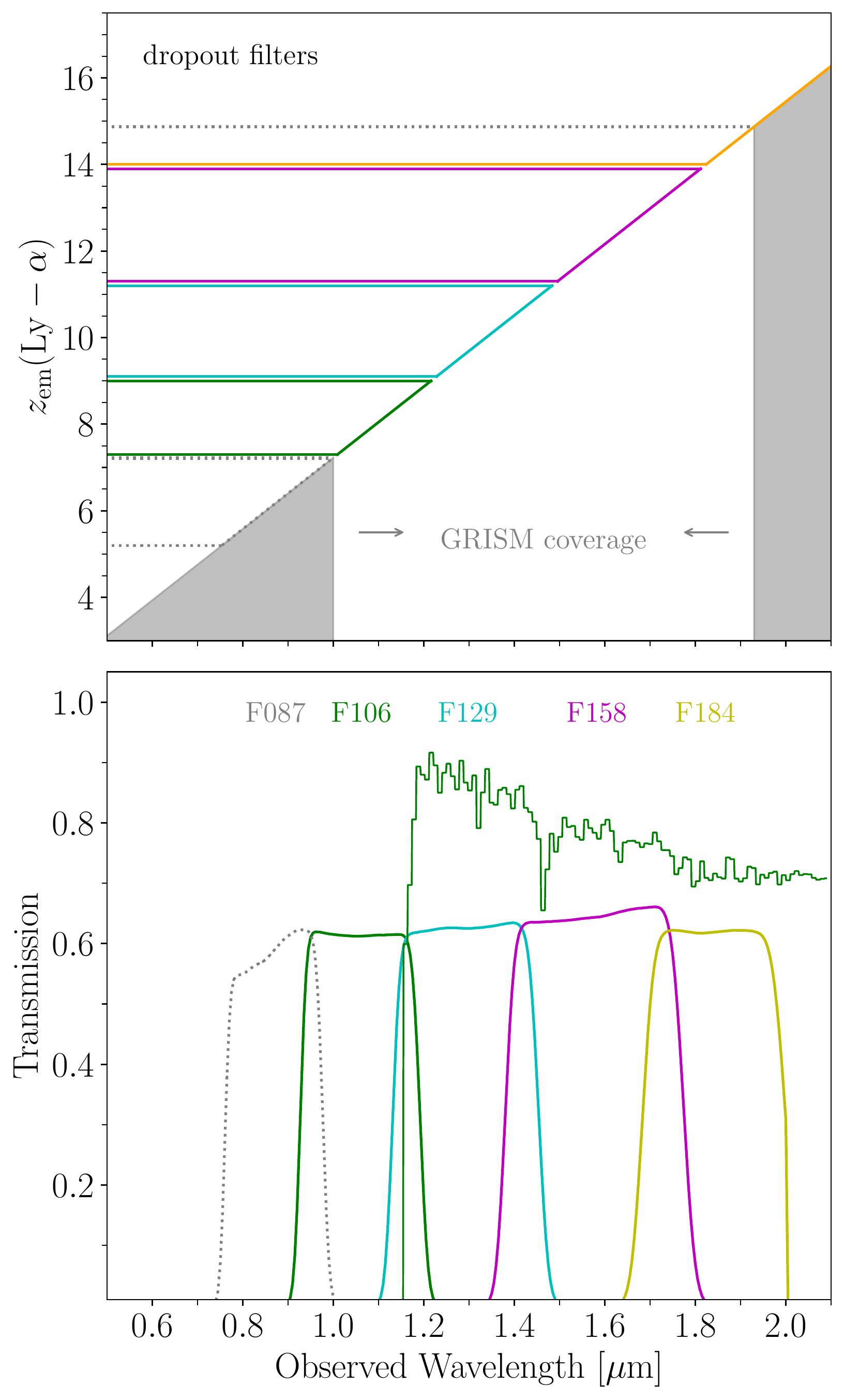}
  \caption{\textbf{Roman spectral coverage.} \textit{Bottom:} we show transmission curves for five filters, starting with the z band (F087), which is not used in the nominal HLS, followed by the Y, J, H, and F184 band filters, from left to right, which will all have HLS coverage. An example spectrum of a $z=8.5$ dust-free galaxy from our model is shown for reference (with nebular emission neglected for clarity). \textit{Top:} We show the corresponding redshift range probed by each filter, where the ``dropout'' filter is the bluest filter containing the $\Lya$ line. The fact that grism coverage is restricted to $\lambda \geq 1 \ \mu \rm{m}$ limits spectroscopic redshifts to galaxy candidates at $z \gtrsim 7.2$.}
  \label{fig:roman_photometry}
\end{figure}

The first requirement is explored in Figure \ref{fig:roman_photometry}, where we show the spectral coverage of the Roman photometry and spectroscopy. The bottom panel shows the transmission curves for the filters that will be used to find dropouts, while in the top panel, we show the redshift ranges corresponding to dropouts in each filter. We follow \citet{Drakos2022}, and for simplicity assign the dropout filter as the bluest filter which contains $\Lya$. All galaxy magnitudes are reported in the filter just redward of the dropout filter, in order to avoid contamination from the break and the $\Lya$ line itself. Though the dropout technique can be used to identify galaxies at redshifts at $z \lesssim 4$ with Roman, the grism covers only $\lambda \in [1, 1.93] \mu \rm{m}$. As a result, only photometrically-selected galaxies at $z \gtrsim 7.2$ have a chance at a $\Lya$ detection.

The second requirement for a galaxy to be included in cross-correlation analyses---that $\Lya$ is bright enough to be detected---is more difficult to assess. As discussed in Section~\ref{sec:galaxy_models} and shown in Figure~\ref{fig:lya_mab}, given the nominal HLSS line flux limits \citep{Wang2022}, any galaxy detected photometrically should also have detectable $\Lya$ emission \textit{assuming negligible absorption from the IGM}. This is of course an optimistic assumption; In general, the occurrence rate of LAEs will be reduced by intergalactic absorption, particularly sources residing in small bubbles. Though the intrinsic LAE fraction at high redshifts is unknown, simulations suggest that $\Lya$ escape could be non-trivial at the redshifts of interest here \citep{Garel2021,Smith2022}. Furthermore, the HLS preferentially picks out bright sources that very well may live in large bubbles and so be subject to little attenuation from the IGM. We explore $\xLAE=0.1$ as our fiducial case, but also explore contrasting scenarios in which only $1$\% or $100$\% of galaxies detected in the HLIS have detectable $\Lya$ emission in the HLSS. We will denote these cases via $\xLAE=0.01$ and $\xLAE=1$, respectively. The pessimistic scenario may apply if the LAE populations uncovered by the Subaru Hyper-Suprime Cam \citep[HSC; e.g.,][]{Ouchi2018} are representative of all reionization era LAEs, while the $\xLAE=1$ case is included as a maximally optimistic limit.



\newpage

\section{Results}
\label{sec:results}

We first show results for the cross-correlation signal $P_\mathrm{21cm\times gal}$ as a function of $k_\perp$ and $k_\parallel$. This is the raw signal that we are attempting to measure. When computing the cross-spectrum, we construct a light cone of observations that span the full redshift range $6 \leq z \leq 15$. In practice, the Roman grism frequency coverage means that we are not sensitive to galaxies at $z < 7.2$. Furthermore, the most detectable contributions to the cross-spectrum will come from relatively low redshift values ($z \lesssim 9$). Nevertheless, we include both higher and lower redshift ranges when constructing the lightcones and then restrict ourselves to $7.5 \leq z \leq 12$ for analysis. Once this has been done, we compute the Fourier transform of non-overlapping windows along the line of sight that each cover 6 MHz of bandwidth. This is representative of the typical bandwidth used in previous HERA data analyses \citep{h1c_limits,h1c_theory}. This bandwidth also translates to comoving distances of $50 \lesssim \chi \lesssim 100$ $h^{-1}$Mpc depending on the redshift. Although these cubes are not strictly comoving, the evolution induced by the light cone is sufficiently small so as not to induce spurious signal \citep{laplante_etal2014}. To further decrease the amount of light cone evolution signal included, we apodize in the line-of-sight direction using a 4-term Blackman-Harris window. We run 30 independent realizations of these simulations where we change the initial conditions but keep the cosmological and astrophysical parameters fixed, and average together the resulting power spectra computed in the described fashion. This averaging helps ensure that the spectra are smooth, especially on the scales relevant to upcoming observations.

\begin{figure}[t]
  \centering
  \includegraphics[width=0.48\textwidth]{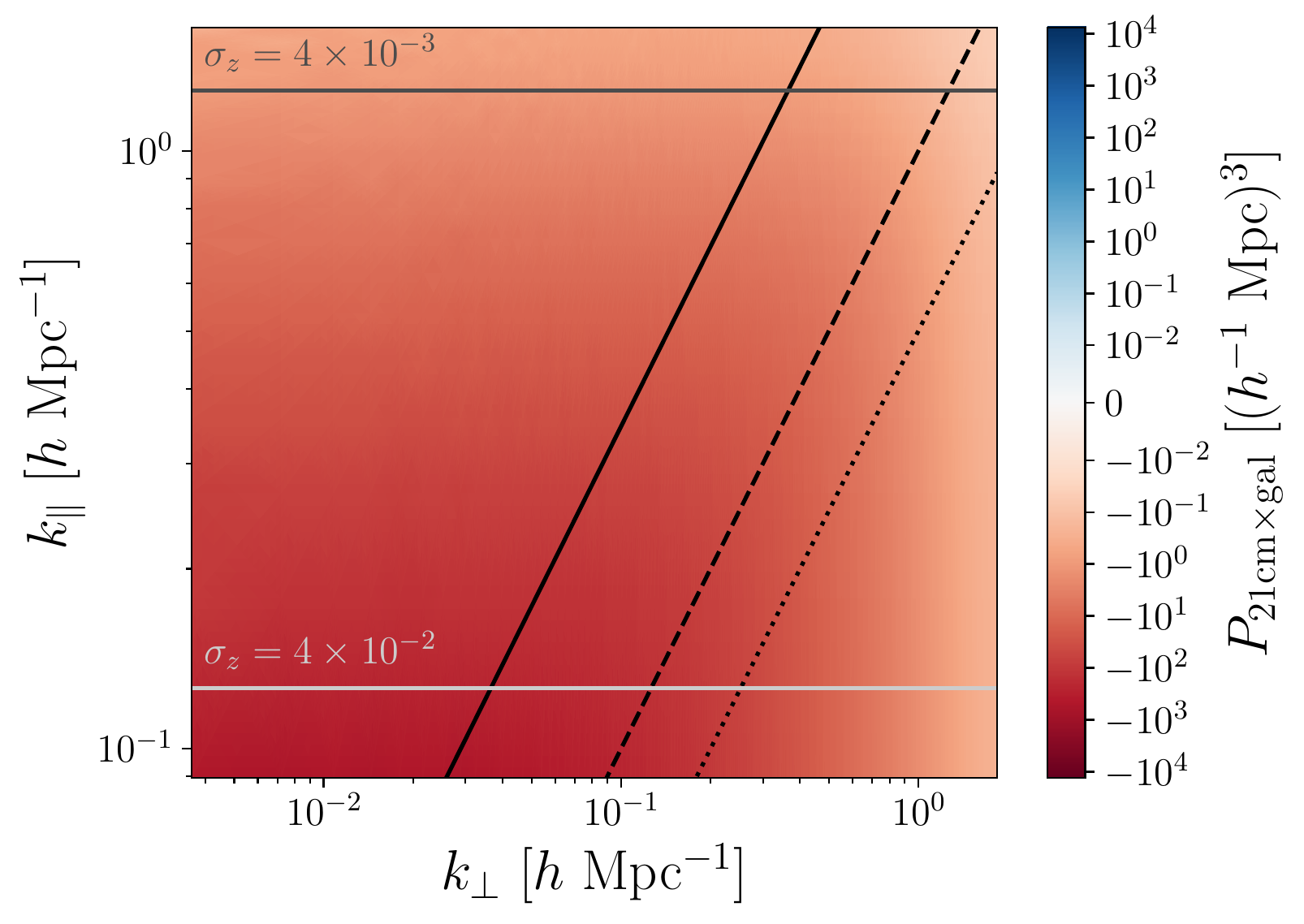}
  \caption{\textbf{The cross-spectrum $P_\mathrm{21cm\times gal}$ as a function of $k_\perp$ and $k_\parallel$.} The spectrum has been computed for a window centered on $z = 8.01$, which is near the midpoint of reionization for our fiducial scenario. The quantity $T_0(z)$ from Equation~(\ref{eqn:t0}) has been divided out. We also plot the slope of the ``wedge'' to demonstrate the extent of foreground contamination. The solid line is the horizon wedge defined by Equation~(\ref{eqn:slope}), and more optimistic values of $m=1$ (dashed) and $m=0.5$ (dotted). Horizontal lines correspond to $k_\parallel$ values associated with uncertainties in redshift determination $\sigma_z$ defined in Equation~(\ref{eqn:gal_noise}).}
  \label{fig:xspec}
\end{figure}

Figure~\ref{fig:xspec} shows the cross-spectrum $P_\mathrm{21cm\times gal}$ near the midpoint of reionization at $z \sim 8.01$, where the amplitude of the cross-spectrum is greatest. In general, the signal is negative, which indicates an anti-correlation. This anti-correlation is expected, as the brightness temperature $\delta T_b$ in Equation~(\ref{eqn:tb}) is sourced by regions of neutral hydrogen, which tend to be low-density regions far from galaxies in an inside-out reionization scenario. We also note that the largest amplitude signal comes from large scales (small values of $k_\perp$ and $k_\parallel$). 
We also show as a solid line the value of the slope of the ``horizon wedge'' defined in Equation~(\ref{eqn:slope}). For comparison, we also show values of $m = 1$ (dashed) and $m = 0.5$ (dotted) lines. These represent varying levels of foreground contamination: the horizon wedge reflects data that are maximally corrupted, whereas less severe cuts would be possible if improved calibration methods make it possible to recover these data. It is also worth noting that there is a much larger dynamic range in $k_\perp$ than in $k_\parallel$: given the fact that we perform a Fourier transform on a slab with a relatively short axis along the line of sight, there are far fewer $k_\parallel$ modes available for a given observation. In order to provide a more significant cumulative measurement of the signal, we combine measurements from non-overlapping redshift windows along the line of sight. Although precise characterization of the $k$- and $z$-dependence of the signal would be ideal, for the near-term forecast most relevant to HERA we are focused on a detection. We also ignore potential covariance between different $k$- and $z$- bins on the assumption that they are relatively free of systematic errors. We return to these assumptions in Section~\ref{sec:21cm_systematics}.

We also show as horizontal lines several values of $k_\parallel$ corresponding to $\sigma_z$ galaxy uncertainties defined in Equation~(\ref{eqn:gal_noise}). We plot values where $k_\parallel = 1/\sigma_\chi$, above which the galaxy uncertainty exponentially increases. Given the exponential nature of the noise contributed from this uncertainty, values of $k_\parallel$ greater than these lines face significant contamination. That is, modes with larger $k_\parallel$ are poorly measured in the galaxy survey and do not contribute significantly to a cross-spectrum detection (see further discussion in Section~\ref{sec:detectability}).
As mentioned above in Section~\ref{sec:galaxy_surveys}, this implies that photometric surveys are insufficient for present purposes, since the redshift uncertainties in high-redshift Lyman-break surveys are expected to be on the order of $\sigma_z \simeq 0.5$.

\section{Detectability}
\label{sec:detectability}

We now turn to forecast the expected signal-to-noise ratio for future cross-spectrum measurements. 
In this analysis, we primarily concern ourselves with the statistical uncertainty of these measurements, and follow related calculations in \citet{furlanetto_lidz2007} and \citet{lidz_etal2009}. For 21\,cm experiments, we examine the projected sensitivity for HERA. For galaxy surveys, we concern ourselves primarily with the Roman HLS and HLS-like surveys. We do not consider the effect of systematic errors for two reasons: (i) in regions of Fourier space that are not dominated by foreground contamination of the wedge, the Fourier modes should be noise dominated, and (ii) in general cross-correlation measurements are unbiased (at the cost of higher statistical variance) if there are no sources of joint systematic error between the two measurements. 
We begin with a more detailed description of the observational approaches of Roman and HERA in Section~\ref{sec:experiments}, go through a treatment of the S/N in our fiducial reionization scenarios in Section~\ref{sec:s/n}, explore building up cumulative S/N by combining measurements from different modes in Section~\ref{sec:cum_snr}, and then turn to possible systematic errors in Section~\ref{sec:systematics}.

\begin{figure*}[t]
  \centering
  \includegraphics[width=0.48\textwidth]{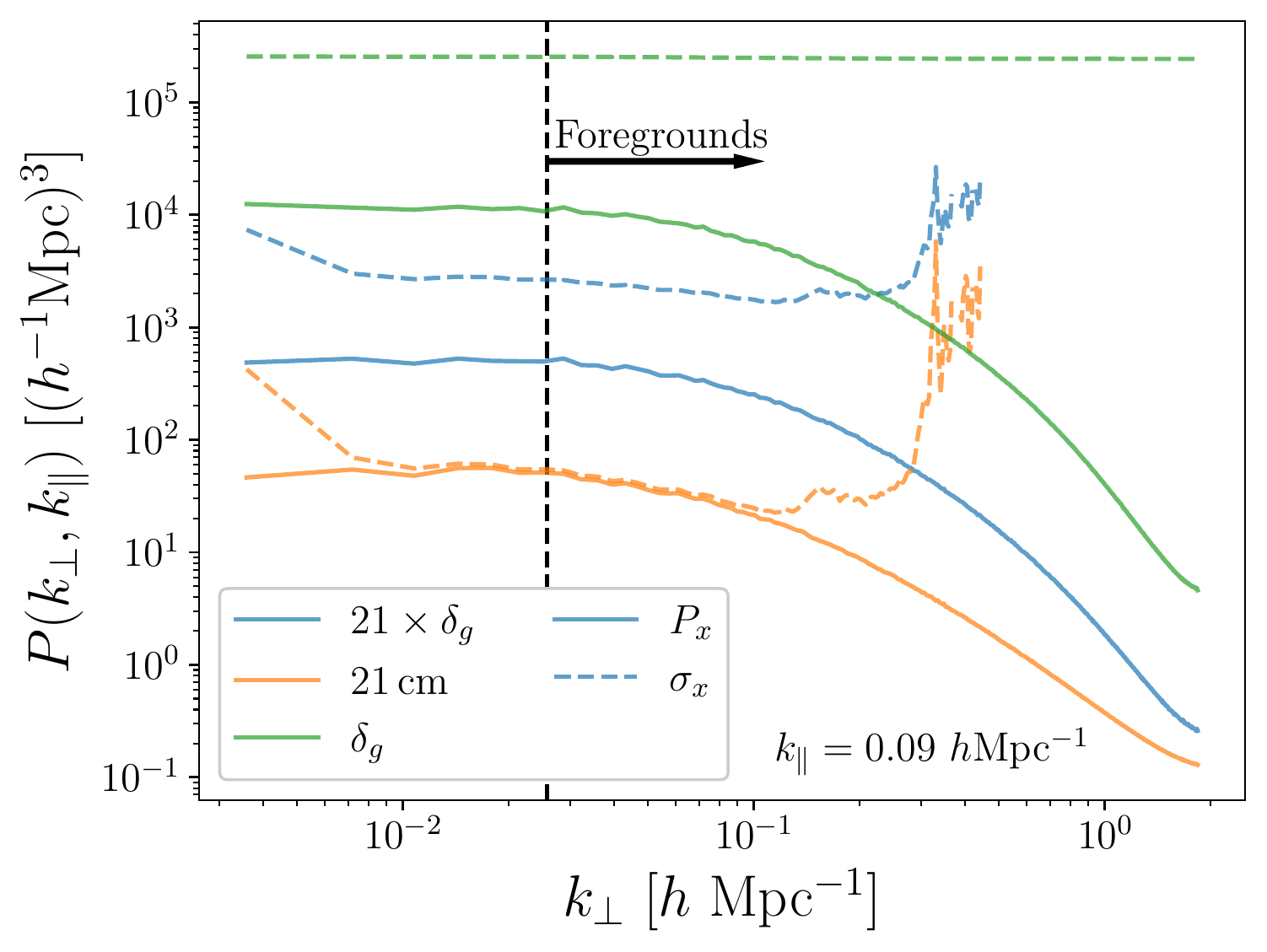}\hfill%
  \includegraphics[width=0.48\textwidth]{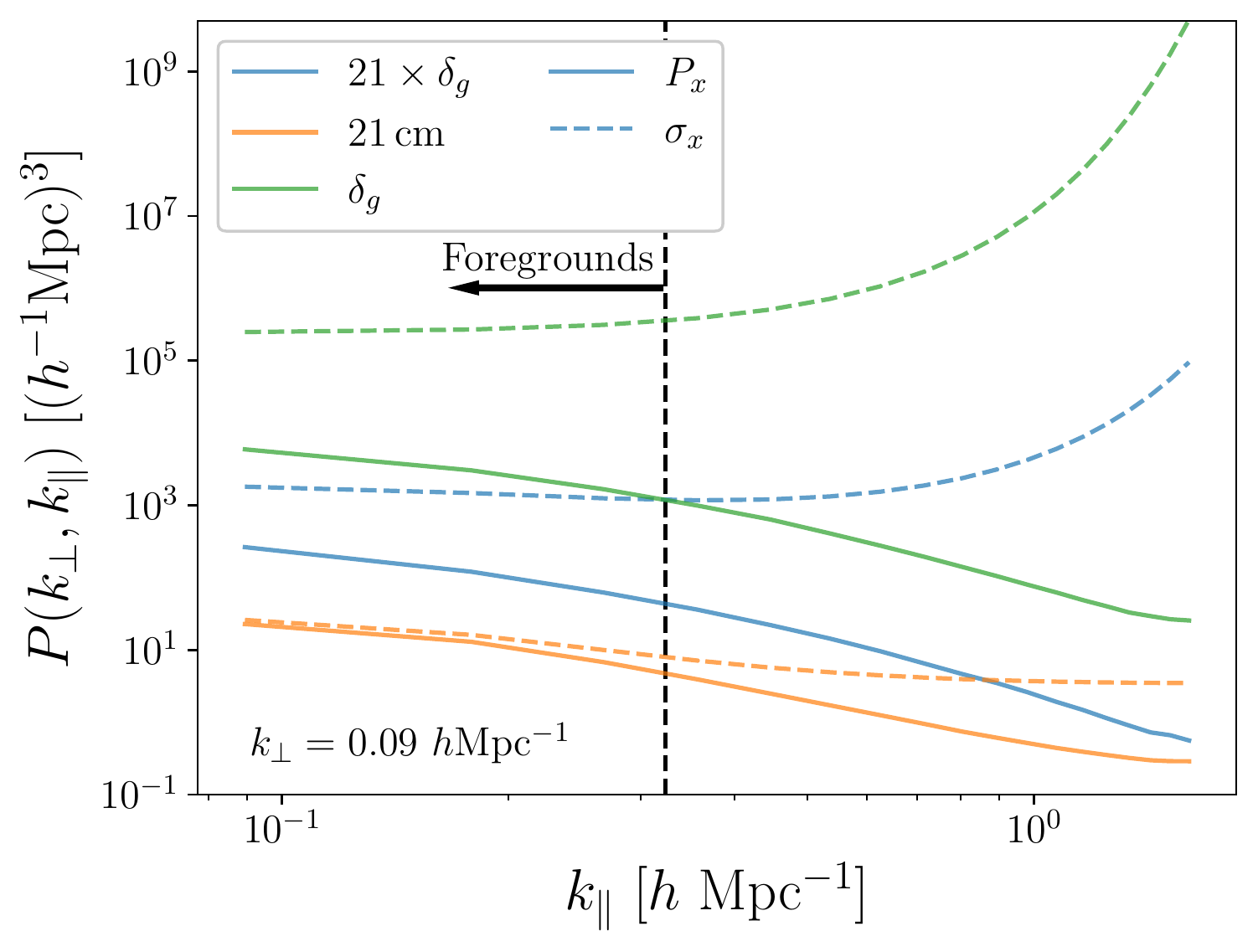}
  \caption{\textbf{The signal and noise quantities of the various components of Equation~(\ref{eqn:xcorr_noise}).} Solid lines show the signal component, and dashed lines show the corresponding noise. Note that the cross-spectrum signal $P_\mathrm{21\times \delta_g}$ is negative, so we have plotted the absolute value. Left: components as a function of $k_\perp$ at a fixed value of $k_\parallel$. Right: components as a function of $k_\parallel$ at a fixed value of $k_\perp$. Vertical dashed lines show the location of the horizon wedge given the fixed values of $k_\parallel$ (left) and $k_\perp$ (right), where the arrows point toward regions that are dominated by foreground contamination.}
  \label{fig:snr_components}
\end{figure*}

\subsection{S/N Calculations}
\label{sec:s/n}

We are interested in the uncertainty of the cross-spectrum $\sigma_\mathrm{21,gal}$ as a function of $k_\perp$ and $k_\parallel$, as this defines our expected sensitivity in Fourier space. In this case, the uncertainty can be expressed as \citep{lidz_etal2009}:
\begin{align}
\sigma_\mathrm{21,gal}^2 &= \mathrm{var}\qty[\frac{1}{T_0(z)}P_\mathrm{21\times gal}(k_\perp, k_\parallel)] \notag \\[0.5em]
&= \frac{1}{2} \Big[P_\mathrm{21\times gal}^2(k_\perp, k_\parallel)  \notag \\
&\quad + \sigma_{21}(k_\perp, k_\parallel) \sigma_\mathrm{gal} (k_\perp, k_\parallel)\Big]. \label{eqn:xcorr_noise}
\end{align}
Similarly, the uncertainty of the 21\,cm power spectrum $\sigma_{21}$ can be written as:
\begin{align}
\sigma_{21}^2 &= \mathrm{var}\qty[P_{21}(k_\perp, k_\parallel)] \notag \\
&= \qty[\frac{1}{T_0(z)^2} P_{21}(k_\perp, k_\parallel) + \frac{1}{T_0(z)^2} P_{21}^\mathrm{noise}(k_\perp, k_\parallel)]^2,
\label{eqn:21cm_tot_noise}
\end{align}
where $T_0$ is defined in Equation~(\ref{eqn:t0}) and $P_{21}^\mathrm{noise}$ is defined in Equation~(\ref{eqn:21cm_noise}). Finally, the uncertainty of the galaxy power spectrum $\sigma_\mathrm{gal}$ can be written as:
\begin{align}
\sigma_\mathrm{gal}^2 &= \mathrm{var}\qty[P_\mathrm{gal}(k_\perp, k_\parallel)] \notag \\
&= \qty[P_\mathrm{gal}(k_\perp, k_\parallel) + P_\mathrm{gal}^\mathrm{noise}(k_\parallel)]^2,
\label{eqn:gal_tot_noise}
\end{align}
where $P_\mathrm{gal}^\mathrm{noise}$ is defined in Equation~(\ref{eqn:gal_noise}). These expressions capture the total variance of the observed power spectra, which is due both to cosmic variance and statistical instrumental uncertainties.

Figure~\ref{fig:snr_components} shows the signal and noise quantities defined in Equations~(\ref{eqn:xcorr_noise}-\ref{eqn:gal_tot_noise}). The solid lines correspond to the signal, and the dashed lines correspond to the noise. To build intuition for how these quantities evolve with Fourier modes $k_\perp$ and $k_\parallel$, we hold one mode fixed and plot the signal and noise curves as a function of the other one. We show results for $f_\mathrm{LAE} = 0.1$ and $\sigma_z = 0.01$ with a redshift window centered at $z = 8.01$, near the midpoint of reionization. The panel on the left shows the behavior for fixed $k_\parallel = 0.09$ $h$Mpc$^{-1}$, and varies $k_\perp$. Also shown as a vertical dashed line is the location of the horizon wedge given the value of $k_\parallel$. As denoted in the figure, the modes to the right of this line are nominally contaminated by foregrounds with a horizon-level wedge defined in Equation~(\ref{eqn:slope}). In this plot, the effect of the $uv$ coverage of HERA can be seen in the 21\,cm noise curves: at very low values ($k_\perp \lesssim 0.01$ $h$Mpc$^{-1}$) and high values ($k_\perp \gtrsim 0.3$ $h$Mpc$^{-1}$), the experimental uncertainty increases dramatically because of the finite baseline size and the lack of very long baselines, respectively. As seen in Figure~\ref{fig:nbl_dist}, many of HERA's baselines are relatively short, meaning there is good coverage of intermediate $k_\perp$ modes at the expense of large ones. On the right panel, we show the behavior for fixed $k_\perp = 0.09$ $h$Mpc$^{-1}$ and vary $k_\parallel$. At relatively large values of $k_\parallel$, the noise dominates significantly for the galaxy signal. It is also worth noting that in this panel, the foreground-contaminated modes lie to the left of the vertical dashed line. 

It is worth discussing some of the large-scale behavior of the trends in this figure to understand how the overall S/N ratio is affected by various observational and experimental considerations. First, it is worth noting that for many values of $k_\perp$, the 21\,cm signal is cosmic-variance dominated. Accordingly, for modes that are well-sampled by HERA, a significant detection of the auto spectrum should be possible. Conversely, the galaxy power spectrum as measured by Roman is dominated by experimental uncertainties, especially at large values of $k_\perp$ and $k_\parallel$. For large values of $k_\perp$ there is a drop-off in the amplitude of the signal at these modes, whereas the low sensitivity for large values of $k_\parallel$ is driven primarily by the exponential scaling of the redshift-space uncertainty in Equation~(\ref{eqn:gal_noise}). Note that the variance decreases with the number of modes in a survey, as given by Equation~(\ref{eqn:nmodes}). Thus it is possible to decrease the instrument uncertainty by using a sufficiently large volume. We return to prospects for measuring the galaxy auto-spectrum below in Section~\ref{sec:cum_snr}. Given the uncertainty of the cross-spectrum is essentially the geometric mean between the individual uncertainties, the S/N of the cross-spectrum is not cosmic variance dominated, but also is not as noise-dominated as the galaxy spectrum.


\newpage

\subsection{Observational Strategies}
\label{sec:experiments}

In addition to the per-mode uncertainty described above, another important ingredient in the ultimate sensitivity calculations is the number of Fourier modes that may be observed given our survey volume $V_\mathrm{survey}$. In practice, we consider cross-spectrum estimates in bins in $k_\perp$ and $k_\parallel$. A given bin will generally receive contributions from many independent Fourier modes, with the Fourier-space resolution set by the survey dimensions, and averaging over these modes will decrease the uncertainties in the bin-averaged measurements.

Consider a cylindrical shell/bin of fixed logarithmic extent $\dd{\ln k_\perp}$ and $\dd{\ln k_\parallel}$, where we have constructed an annulus in the plane of the sky with radius $k_\perp$ and at coordinate $k_\parallel$ along the line-of-sight dimension. The number of Fourier modes in our survey that lie within this cylindrical bin is given by:
\begin{equation}
  \dd{N}(k_\perp, k_\parallel) = \frac{k_\perp^2 k_\parallel V_\mathrm{survey}}{(2\pi)^2} \dd{\ln k_\perp} \dd{\ln k_\parallel}.
  \label{eqn:nmodes}
\end{equation}
Given that the underlying fields are real, there is a Hermitian symmetry of the Fourier transform applied when constructing the power spectrum. As such, we restrict ourselves to estimating the variance only from the upper half-plane so as not to double-count these modes when estimating the variance. We have also taken the absolute value of $k_\parallel$ assuming that positive and negative values are statistically similar.\footnote{If one uses linear bins of $\dd{k_\perp}$ and $\dd{k_\parallel}$ instead of logarithmic ones, the number of modes is given by:
\begin{equation*}
\dd{N}(k_\perp, k_\parallel) = \frac{k_\perp V_\mathrm{survey}}{(2\pi)^2} \dd{k_\perp} \dd{k_\parallel},
\end{equation*}
where we have applied similar considerations for mode counting only in the positive half-plane and $k_\parallel$ symmetry.
}

As discussed above in Section~\ref{sec:obs} and shown in Figure~\ref{fig:survey_footprints}, we expect that the overlap between Roman and HERA is about 500 deg$^2$. Given that HERA is a drift-scan instrument (rather than a tracking/pointing array), we can treat a joint measurement as a series of single-field patches that we sum together incoherently. This approach is similar to the one advocated in \citet{liu_shaw2020}, where the number of such patches is parameterized as $N_\mathrm{patch}$. The instantaneous field-of-view (FoV) of HERA is about 10 deg \citep{deboer_etal2017}, so we take the number of such patches to be $N_\mathrm{patch} = 5$. We divide by the square root of this number when estimating the uncertainties in different wavenumber bins.

As a trade-off to this patch-based approach, we must compute our survey volume $V_\mathrm{survey}$ self-consistently given the angular extent of one of these patches. For each redshift bin $z_i$, we assume the FoV of HERA and use this value to compute the comoving distance $\chi_\perp$ using the angular diameter distance $D_A(z_i)$. For the line-of-sight component, we compute the comoving distance $\chi_\parallel$ that corresponds to our 6 MHz of bandwidth. The resulting survey volume is thus $V_\mathrm{survey} = \chi_\perp^2 \chi_\parallel$. In practice the number of non-overlapping patches will be dictated by the joint coverage of HERA and Roman. For the forecasts considered here, we assume that it is possible to construct $N_\mathrm{patch}=5$ fields of observation from which we estimate our S/N. The overall uncertainty is only sensitive to the square root of this quantity, and so having fewer patches does not significantly affect the overall forecast. 

\subsection{Combining Modes}
\label{sec:cum_snr}

As discussed above in Section~\ref{sec:experiments}, the number of times an individual mode is measured depends on the details of the survey geometry. To first order, we have approximated this counting factor using Equation~(\ref{eqn:nmodes}), which gives the differential number of modes within a volume element of $(k_\perp, k_\parallel)$ space. By accounting for this factor, we can understand how the expected variance of the measurement is affected by the survey volume and the amount of sky area covered by the two different experiments. We define the bin-averaged S/N $\hat{s}$ as:
\begin{equation}
\hat{s}(k_\perp, k_\parallel) \equiv \sqrt{N_\mathrm{patch}\dd{N}(k_\perp,k_\parallel)}\frac{P_\mathrm{21\times gal}(k_\perp, k_\parallel)}{\sigma_\mathrm{21\times gal}(k_\perp, k_\parallel)},
\label{eqn:s-hat}
\end{equation}
where $P_\mathrm{21\times gal}$ is measured empirically from our simulated volumes and $\sigma_\mathrm{21\times gal}$ is computed from Equation~(\ref{eqn:xcorr_noise}). For the forecasts here, we choose logarithmic bins of fixed extent $\dd{\ln k_\perp} = \dd{\ln k_\parallel} = 0.1$, though we have verified that the overall results are relatively insensitive to the precise choice of binning scheme.

\begin{figure}[t]
  \centering
  \includegraphics[width=0.48\textwidth]{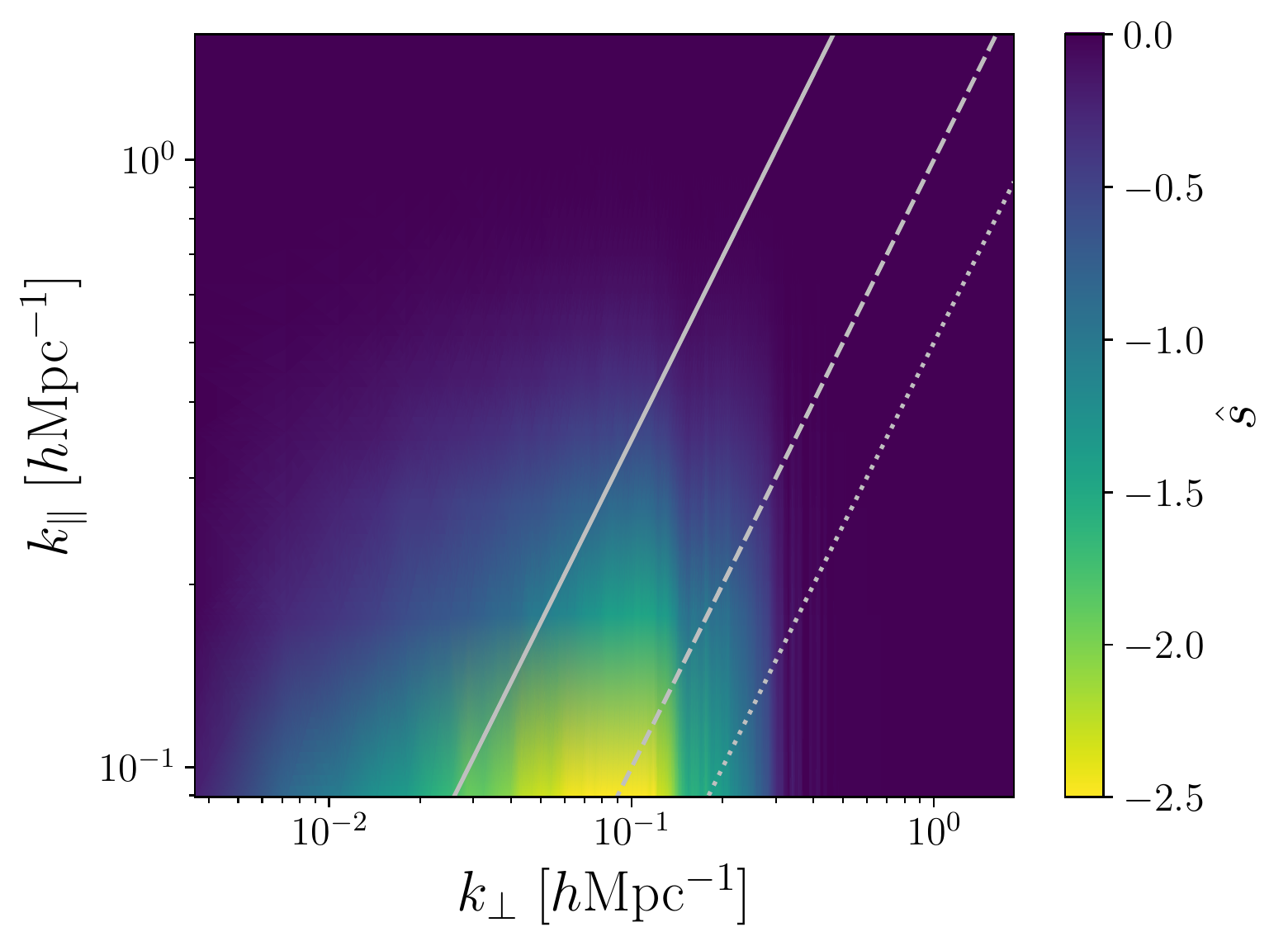}
  \caption{\textbf{The S/N ratio as a function of $k_\perp$ and $k_\parallel$ of the $P_\mathrm{21,gal}$ cross-spectrum, with the uncertainty $\sigma_\mathrm{21,gal}$ given by Equation~(\ref{eqn:xcorr_noise}).} As in Figure~\ref{fig:xspec}, we show the horizon-wedge (solid), $m=1$ (dashed) and $m=0.5$ (dotted) foreground contamination lines. The plotted noise levels assume a photometric redshift uncertainty of $\sigma_z = 0.01$ and an effective LAE observation fraction of $f_\mathrm{LAE} = 0.1$. For several modes the S/N is greater than 1, and combining the individual uncertainties can lead to even more significant detection.}
  \label{fig:snr}
\end{figure}

Figure~\ref{fig:snr} shows the per-mode S/N ratio as a function of $k_\perp$ and $k_\parallel$ of the cross-spectrum $P_\mathrm{21,gal}$ for our fiducial reionization scenario. The uncertainty $\sigma_\mathrm{21\times gal}$ is given by Equation~(\ref{eqn:xcorr_noise}), and the signal $P_\mathrm{21\times gal}$ is shown in Figure~\ref{fig:xspec}. For the uncertainty calculation, we use the parameters $f_\mathrm{LAE} = 0.1$ and $\sigma_z = 0.01$. When the mode-sampling in Equation~(\ref{eqn:nmodes}) is applied, we see that the the significance of several modes is $\hat{s} > 1$. Also consistent with Figure~\ref{fig:snr_components}, the most significant modes lie in the regions of $k$-space that have small values of $k_\perp$ and $k_\parallel$. We also show the values of wedge corresponding to the horizon (solid), $m=1$ (dashed) and $m=0.5$ (dotted). In principle there is a significant amount of sensitivity that can be obtained by increasing the number of foreground modes that can be used, though as shown below a significant detection is possible even if restricted to modes that are expected to be free of foreground contamination. 

With these individual uncertainties defined in Equation~(\ref{eqn:s-hat}), we compute the cumulative S/N by adding the per-bin sensitivity $\hat{s}$ for individual $(k_\perp, k_\parallel)$ bins in quadrature:
\begin{equation}
\qty(\frac{S}{N})^2 = \sum_{k_\perp, k_\parallel, z} \hat{s}^2(k_\perp, k_\parallel, z).
\label{eqn:snr}
\end{equation}
In this way, we can build up cumulative sensitivity by combining measurements from different modes together. We also sum over all non-overlapping redshift windows between $7.2 \leq z \leq 12$. This restriction on the redshift value reflects the inability of the Roman grism to observe and detect LAEs below this threshold. In practice, given the low number density of galaxies expected above $z \gtrsim 10$, the sensitivity of these measurements decreases significantly towards high redshift.

To capture how the total uncertainty varies with changes made to individual experimental parameters, we explore a  multi-dimensional parameter space of statistical errors.

Next, we consider how the error bars on the cross-power spectrum vary with changes in both the survey and model parameters.  
Specifically, we explore how the S/N varies with respect to: (i) the slope of the 21\,cm wedge (i.e., the degree to which foregrounds contaminate the signal of interest), (ii) the galaxy redshift uncertainties, $\sigma_z$, (iii) the effective duty cycle of sources, $f_\mathrm{LAE}$, (iv) whether the spin-temperature of the IGM is saturated or maximally cold, (v) the ionization history, and (vi) the galaxy bias factor predicted by \textsc{ares} and \textsc{BlueTides}. Exploring these variations will help in formulating optimal observing strategies for upcoming cross-power spectrum measurements.


\begin{deluxetable*}{@{\extracolsep{6pt}}ccccccc}
\tablecaption{Cumulative S/N Ratio for Different Model Parameters
\tablewidth{0pc}
\label{table:snr}}
\tablehead{\colhead{Bias Model} & \colhead{Ionization History} & \colhead{IGM} & \colhead{$f_\mathrm{LAE}$} & \colhead{$\sigma_z$} & \colhead{Wedge Cut} & \colhead{S/N}}
\startdata
\textbf{\textsc{ares}} & \textbf{fiducial} & \textbf{saturated} & \textbf{0.1} & \textbf{0.01} & \textbf{Horizon} & \textbf{12} \\
\cline{6-7}
\textsc{ares} & fiducial & saturated & 0.1 & 0.01 & 1.0 & 28 \\
\textsc{ares} & fiducial & saturated & 0.1 & 0.01 & 0.5 & 32 \\
\cline{5-7}
\textsc{ares} & fiducial & saturated & 0.1 & 0.001 & Horizon & 14 \\
\textsc{ares} & fiducial & saturated & 0.1 & 0.1 & Horizon & 0.55 \\
\cline{4-7}
\textsc{ares} & fiducial & saturated & 1 & 0.01 & Horizon & 35 \\
\textsc{ares} & fiducial & saturated & 0.01 & 0.01 & Horizon & 3.7 \\
\cline{3-7}
\textsc{ares} & fiducial & cold & 0.1 & 0.01 & Horizon & 15 \\
\cline{2-7}
\textsc{ares} & short & saturated & 0.1 & 0.01 & Horizon & 10.0 \\
\textsc{ares} & late & saturated & 0.1 & 0.01 & Horizon & 9.5 \\
\hline
\textsc{BlueTides} & fiducial & saturated & 0.1 & 0.01 & Horizon & 29 \\
\textsc{BlueTides} & fiducial & saturated & 0.1 & 0.01 & 1.0 & 69 \\
\textsc{BlueTides} & fiducial & saturated & 0.1 & 0.01 & 0.5 & 81 \\
\enddata
\tablecomments{The cumulative S/N ratio for different combinations of model parameters and observational assumptions. Our fiducial parameters assume:
the reionization history described in Section~\ref{sec:ion_hist}, a spin-temperature saturated IGM, $f_\mathrm{LAE} = 0.1$, $\sigma_z = 0.01$, and a horizon wedge level of foreground contamination. We show how the resulting S/N changes when modifying each of these parameters around these fiducial values.}
\end{deluxetable*}

Table~\ref{table:snr} shows the cumulative S/N calculated using Equation~(\ref{eqn:snr}) for our various combinations of parameters. For our default parameters, we choose a horizon wedge level of foreground contamination, $f_\mathrm{LAE} = 0.1$, $\sigma_z = 0.01$,\footnote{This value of $\sigma_z$ is consistent with the redshift-dependent scaling of $\sigma_z(z) = 0.001(1+z)$ mentioned above \citep{Wang2022}.}, a spin-temperature saturated IGM, our fiducial reionization scenario, and the galaxy bias $b_g(z)$ as predicted by \textsc{ares}. We show the resulting changes to the S/N as a function of varying these parameters. Encouragingly, given these default parameters, we forecast a 12$\sigma$ detection for the nominal HERA and Roman sensitivities. For the same combination of observational parameters, we forecast a 282$\sigma$ detection of the 21\,cm auto power spectrum from HERA and a 14$\sigma$ detection of the LAE galaxy auto power spectrum from Roman.\footnote{Note that for the HERA and Roman auto spectra, we have assumed sky coverage areas of 1000 deg$^2$ and 2200 deg$^2$ respectively. Also note that the LAE galaxy auto power spectrum does not include the foreground wedge excision, so the nominal S/N forecast is comparable to the cross-power spectrum.} We discuss further practical challenges in achieving this cross-power S/N in Section~\ref{sec:systematics}. 

We now briefly discuss some of the general trends on display in Table~\ref{table:snr} to contextualize some of the various observational effects at play. Not surprisingly, decreasing the slope of the foreground wedge significantly increases the cumulative S/N ratio. If a value of $m = 0.5$ is assumed, the cumulative S/N increases by nearly a factor of 3, to a roughly 32$\sigma$ detection. As shown in Figure~\ref{fig:snr}, there are a non-trivial number of modes that are subject to foreground contamination, which are also the ones most strongly detectable. By increasing the number of modes that contribute to the S/N statistic, the prospects for successful detection and characterization are improved dramatically. Furthermore, this improvement relies almost exclusively on improvements to 21\,cm calibration and analysis techniques. Although the design of HERA means that reliably accessing these modes may be difficult, it is worth exploring whether there are ways to extract and use some of this information when measuring the cross-correlation spectrum.

For the redshift uncertainty $\sigma_z$, we see that there is asymmetric behavior in how the overall uncertainty changes. Increasing the fiducial value to $\sigma_z = 0.001$ offers a modest improvement to 14$\sigma$. However, decreasing the uncertainty to $\sigma_z = 0.1$ leads to a cumulative S/N well below 1. This can be seen in the behavior of the S/N shown in the right panel of Figure~\ref{fig:snr_components}: as $k_\parallel$ increases, the galaxy noise power (and hence the cross-spectrum variance in the shot-noise dominated limit) increases exponentially. When the uncertainties are such that $\sigma_z \sim 0.1$ (which is the expected level for photometrically determined redshift values), there are very few $k_\parallel$ values that have an appreciable S/N value. It may be possible to access lower values of $k_\parallel$ by increasing the bandwidth used to observe 21\,cm measurements, though the foreground contamination swamps the signal for small values of $k_\parallel$ \citep{h1c_limits}.

The effective LAE duty cycle $f_\mathrm{LAE}$ also has a significant effect on the projected S/N ratio. As discussed above, changing this quantity only affects the galaxy number density calculation when determining the shot-noise term of the galaxy uncertainty $P_\mathrm{gal}^\mathrm{noise}$ in Equation~(\ref{eqn:gal_noise}). Nevertheless, this particular source of uncertainty is one of the most significant. The left panel of Figure~\ref{fig:snr_components} suggests that this shot-noise term dominates the galaxy spectrum uncertainty at all values of $k_\perp$, which in turn limits the sensitivity of the cross-spectrum measurement. Thus, increasing this quantity to $f_\mathrm{LAE} = 1$ significantly reduces the shot-noise uncertainty and increases the S/N by a factor of about 3. Conversely, using $f_\mathrm{LAE} = 0.01$ decreases the S/N by a similar factor.

Next, we look at the impact of a saturated versus a cold IGM. As mentioned above in Section~\ref{sec:zreion}, our default model assumes that the spin-temperature of hydrogen has been saturated, so that $T_s \gg T_\gamma$. However, we also investigate the case of a ``cold'' IGM, in which the gas outside fully ionized regions is not heated. As a result, the 21\,cm brightness temperature can be large and negative, which increases the amplitude of the 21\,cm auto power spectrum as well as the cross-spectrum of interest here. Although this assumption significantly increases the amplitude of the 21\,cm auto power spectrum, it only provides a modest increase in the cross-spectrum here \citep[see also][]{Heneka2020}, largely due to the quadratic scaling of the fluctuation amplitude in the 21\,cm auto power spectrum versus the linear scaling here. As discussed above, the 21\,cm signal is primarily cosmic-variance limited for the detectable Fourier modes in the cross-power spectrum, and so the higher amplitude fluctuations in the cold IGM boost both the signal and the noise and do not significantly improve the total S/N. 

Furthermore, we investigate the effects of changing the ionization history. As can be seen in Table~\ref{table:snr}, both the short and the late reionization scenarios lead to slightly smaller projected S/N values compared to the fiducial history chosen here. For the case of the short history, the cross-power spectrum itself has a slightly larger amplitude due to a larger 21\,cm signal, noted as a feature in previous applications of this seminumeric model \citep{laplante_etal2014}. However, this increase in signal amplitude is offset by having fewer redshift windows over which the 21\,cm and galaxy signals have an appreciable cross-spectrum amplitude. 
For the case of the late reionization scenario, imposing the cutoff of $z \geq 7.2$ means that spectral windows near the midpoint of reionization are thrown out. Despite the smaller amount of redshift information that can be used, there is still enough sensitivity to detect the cross-spectrum at roughly 9.5$\sigma$.

Finally, we also look at the effect of the linearly-biased galaxy model chosen for the galaxy fields. Instead of our models generated from \textsc{ares}, we use bias values and number densities inferred from the \textsc{BlueTides} simulation. As can be seen in Figure~\ref{fig:bias_models}, the galaxy properties predicted by \textsc{BlueTides} feature both a larger galaxy bias $b_g$ and a greater galaxy number density $n_g$. The combination of these effects is a significantly larger predicted value for the cumulative S/N, with a value of 29$\sigma$. In Table~\ref{table:snr}, we also show the S/N for different values of the horizon cut. In general, the S/N for \textsc{BlueTides} is larger than the S/N for \textsc{ares} by a factor of 2-3. We find that this trend holds for all of the parameter combinations we explored.

\subsection{Sources of Systematic Errors}
\label{sec:systematics}

In addition to the statistical uncertainties discussed above, there are sources of systematic errors both for galaxy and 21\,cm uncertainties. We begin by discussing issues related to galaxy observations and potential paths for mitigation, then turn to 21\,cm measurements.

\subsubsection{Galaxy Survey Errors}
\label{sec:galaxy_systematics}
For the galaxy surveys, there is the potential for the measured redshift values from $\Lya$ lines to be systematically offset from the true galaxy redshifts \citep[e.g.,][]{Shapley2003,Erb2014,Shibuya2014,Mason2018,Endsley2022}. 
As we show in the above analysis, accurately determined spectroscopic redshift values are key to yielding a statistically significant detection.

Fortunately, there are two potential solutions to systematic errors in the determination of galaxy redshift values. First, we may use data taken from other instruments as a means of cross-validating and calibrating the measurements from the HLSS. Specifically, the near-infrared spectrograph (NIRSpec; \citealt{nirspec}) aboard the James Webb Space Telescope (JWST) is expected to take high-resolution spectroscopic measurements of many high-redshift galaxies. By comparing several of the higher quality spectra from JWST with a subset of those taken by Roman, it may be possible to model and remove any systematic uncertainties from the full catalog produced by Roman. Additionally, when endeavoring to measure the cross-correlation signal in practice, we can treat the galaxy redshift values as uncertain quantities, or even a single nuisance parameter quantifying the typical systematic redshift offset \citep[as done in, e.g.,][]{CHIME2022}, and marginalize over them in Markov chain Monte Carlo (MCMC)-style analysis. Essentially, the cross-correlation signal is not significant when the incorrect values for the galaxy sample redshifts are used, but the MCMC technique yields a maximal signal as the accuracy is  increased. We can use the measured redshift values to form relatively tight prior distributions for the values used in the analysis, which are made more accurate as the space is sampled.

Another source of uncertainty is the potential confusion of sources as measured by the grism, given that the instrument covers a relatively wide field of view. These sources can either be other nearby high-redshift galaxies, or interloping low-redshift galaxies or stars. Interlopers that are not high-redshift LAEs will increase the effective shot-noise of the measurement. An additional complication is the confusion of two LAE objects in the Roman survey. As discussed above in Section~\ref{sec:galaxy_surveys}, we expect that objects observed in the HLSS will also be sufficiently bright in the HLIS so as to be well-localized. In \citet{satpathy_etal2020}, the authors explicitly model and measure the expected number of overlapping galaxy spectra. (See their Section~3.1.3.) They find that about 7\% of galaxies are expected to overlap. Assuming that the overlap of galaxies is not biased in some way, then this only results in an effective decrease in the number density of observed galaxies, which is accounted for in our $f_\mathrm{LAE}$ parameter. As we show, even with a pessimistic value of $f_\mathrm{LAE} = 0.01$, we expect a significant detection of the cross spectrum. Thus, this potential overlap should not preclude a successful measurement.


\subsubsection{21\,cm Survey Errors}
\label{sec:21cm_systematics}

On the 21\,cm side, there are also considerations of systematic uncertainties. As discussed at great length throughout the rest of this paper, we know that the wedge and 21\,cm foregrounds pose a significant challenge when making measurements and inferences. Although we have chosen a pessimistic value of a horizon cut for our 21\,cm foregrounds, the amount of contamination may necessitate increasing this value even further. Previous analysis has shown that a ``super-horizon buffer'' may be required to fully remove the effect of foreground emission \citep[e.g.,][]{pober_etal2013}, which would lead to a further decrease in the amount of $(k_\perp, k_\parallel)$ space that can be used for this cross-correlation. Furthermore, although it may be possible for HERA to produce data from inside the wedge, the primary mode of HERA data analysis thus far has been explicit foreground avoidance. The development and implementation of new techniques may be required to achieve levels of foreground cleaning required to improve the prospects for this measurement.

\begin{figure*}[t]
  \centering
  \includegraphics[width=0.48\textwidth]{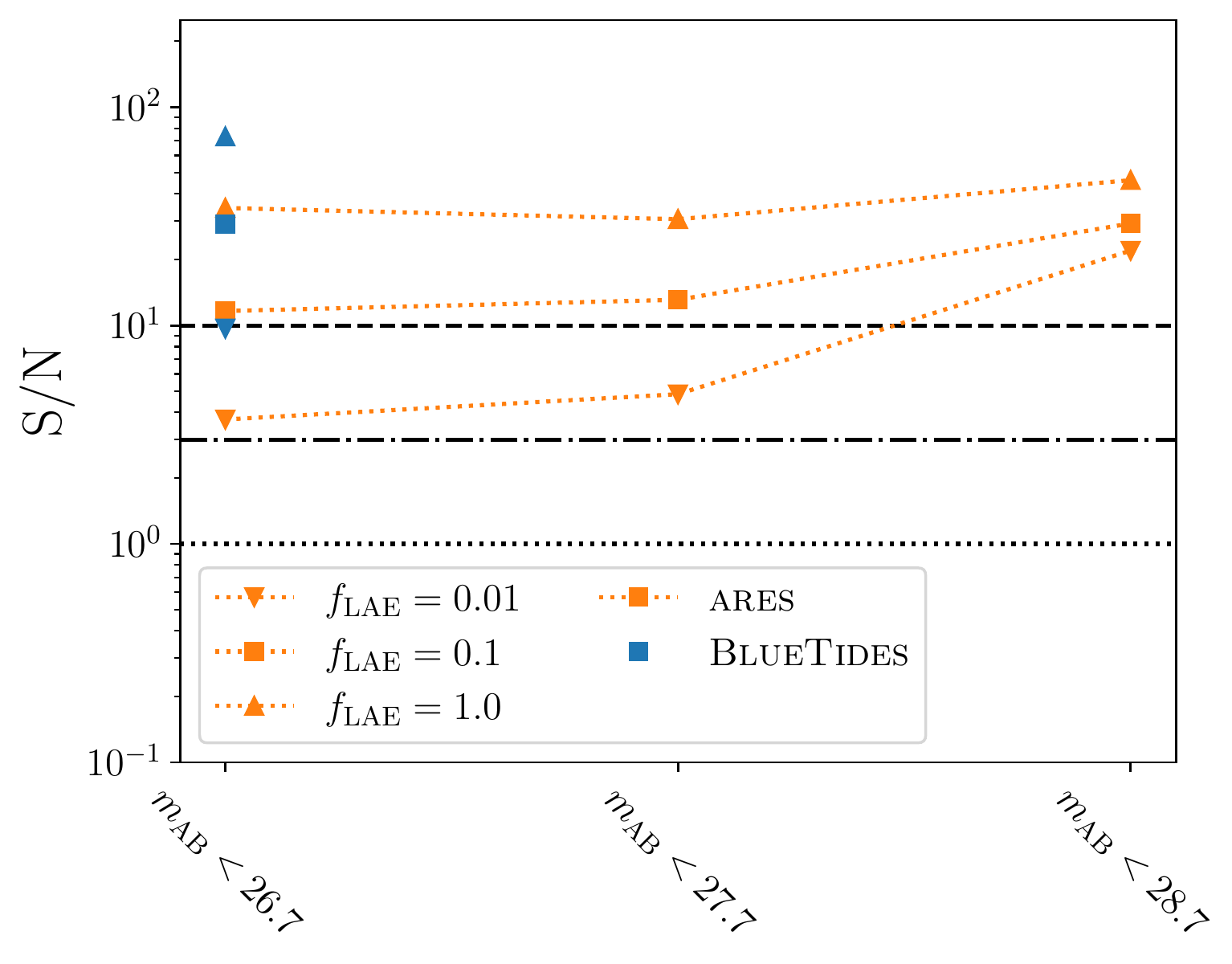}\hfill%
  \includegraphics[width=0.48\textwidth]{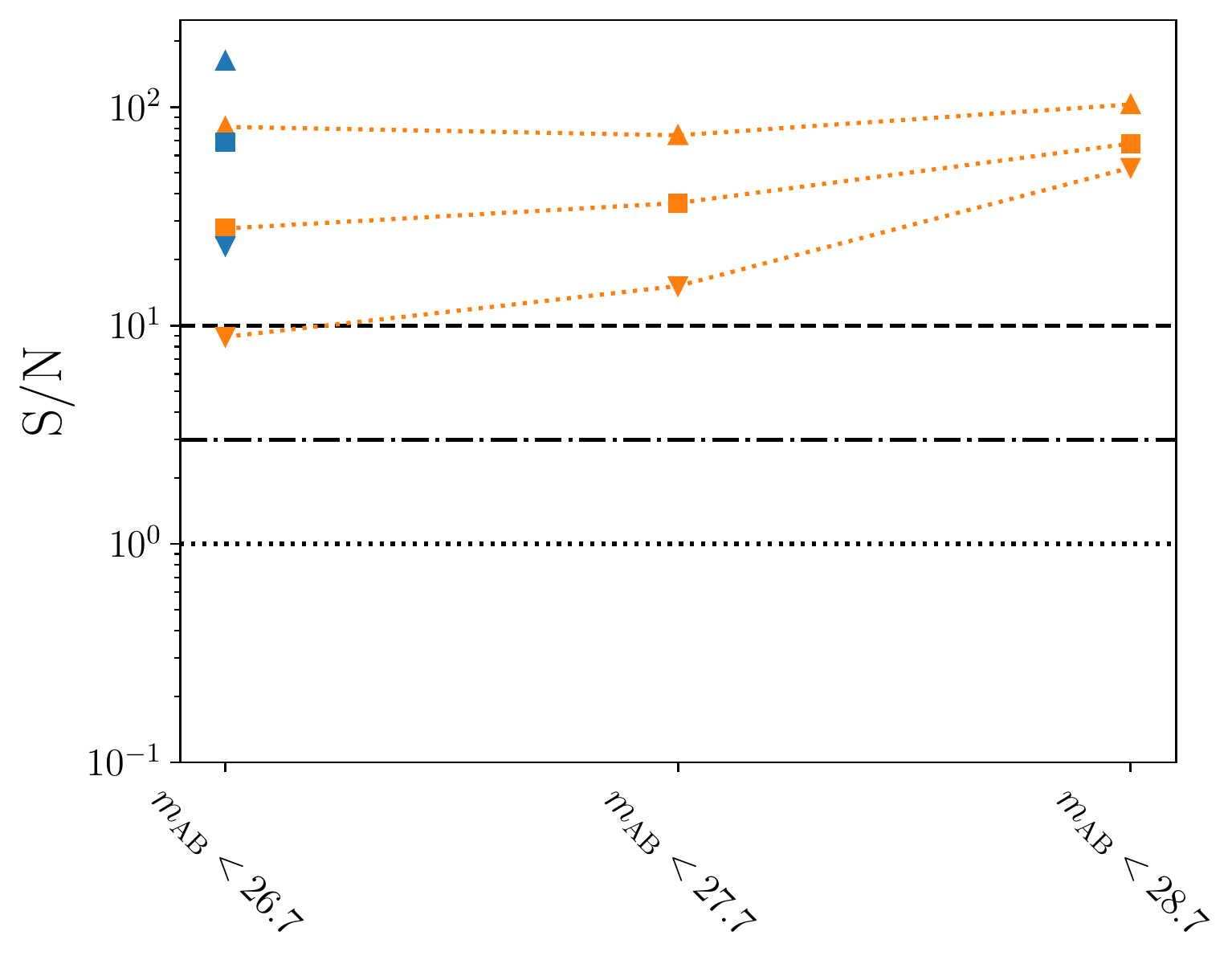}
  \caption{\textbf{The projected S/N as a function of the limiting magnitude $\mab$ for different combinations of observational parameters.} Left: the forecast S/N for a horizon wedge. Right: the S/N for $m=1.0$. We show different values of $f_\mathrm{LAE}$ using different symbols, and use color to denote \textsc{ares} versus \textsc{BlueTides} galaxy models (the latter of which are only shown for $\mab < 26.7$). For our assumed default parameters, one magnitude of deeper observations corresponds to a slight increase of the S/N, from 12$\sigma$ to 13$\sigma$. Note that when considering the deeper observations, we use a sky coverage of 100 deg$^2$ rather than 500 deg$^2$. Going two magnitudes deeper corresponds to a further increase to S/N of 29$\sigma$. See Section~\ref{sec:discussion} for more discussion.}
  \label{fig:deep_fields}
\end{figure*}

Separately to the issue of Fourier modes contaminated by foreground contamination \textit{per se}, we have also ignored contributions of foreground signals to the cross-spectrum variance. Although we expect the foregrounds from the 21\,cm data to be uncorrelated with the galaxy survey signal, they nevertheless contribute to the cross-spectrum variance. As explored above, the variance of the 21\,cm signal contributes to the overall variance of the cross-power spectrum, so increasing the variance will reduce the forecast S/N values.

Another potential issue for 21\,cm measurements is a decrease in the effective number of baselines measuring a given $u$ mode on the sky. Figure~\ref{fig:nbl_dist} shows the distributions of baseline for the full HERA-350 array. As shown in Figure~\ref{fig:snr_components}, this full distribution is expected to yield measurements that are cosmic-variance limited, rather than instrument-noise limited. However, if the effective number of baselines is decreased, this may no longer be true, which would increase the effective noise of the measurement. Fortunately, given the highly redundant nature of HERA, this baseline distribution ensures that these low-$k_\perp$ modes that yield the highest S/N are very well-sampled, and should be cosmic-variance dominated even with a smaller number of antennas.

Another source of systematic errors that may decrease the sensitivity of HERA measurements in practice is the presence of radio frequency interference (RFI), which affects frequency ranges detectable by the instrument. Wider frequency ranges that are relatively free of RFI mean that multiple different redshift windows can be used to increase the cumulative S/N. Additionally, wider frequency ranges mean that broader FFTs can be performed, which probe smaller values of $k_\parallel$. Given that most of the sensitivity for the cross-correlation statistic come from low values of $k_\parallel$, having access to wide frequency ranges free of RFI is appealing. Fortunately, the observed frequency ranges for HERA are relatively wide \citep{h1c_limits}, so it should be possible to construct several frequency windows for evaluating the cross-correlation S/N to make such a detection possible.


\newpage

\section{Discussion}
\label{sec:discussion}

Based on the S/N calculations above in Section~\ref{sec:s/n}, there are several important findings that have implications for various observation strategies. We now turn our attention to the proposed HLS survey, and ask about how to improve prospects for this cross-correlation measurement. Before discussing these prospects in detail, it is worth making some high-level remarks about this trade-off.

As shown above in Section~\ref{sec:s/n} and Figure~\ref{fig:snr_components}, we expect the galaxy shot noise term $1/n_\mathrm{gal}$ to be significantly larger than the intrinsic galaxy power spectrum $P_\mathrm{gal}$ on all scales, but specifically for the low-$k$ modes that contain the most sensitivity in the cross-power spectrum. If $1/n_\mathrm{gal} \gg P_\mathrm{gal}$ on the scales of interest, then one is shot-noise limited, and it is better to attempt to make deeper observations so that the shot-noise term decreases. Conversely, once the observations are sufficiently deep such that $1/n_\mathrm{gal} \sim P_\mathrm{gal}$, then one should use wider observations at this depth rather than deeper ones. We now explore some of the quantitative ramifications for the HERA--Roman cross-spectrum.

\subsection{Increased Sky Coverage}
Let us first suppose that we were able to make use of additional time on the instrument, and decided to cover additional sky area (in the HERA stripe) at the same depth. Using the noise formalism outlined above, the effect of increasing the survey area when cross-correlating with HERA is related to the number of non-overlapping sky patches that are observed $N_\mathrm{patch}$. Specifically, by assuming that the signal in these patches average together incoherently, we increase the computed S/N value by $\sqrt{N_\mathrm{patch}}$. In the above analysis, given that the projected overlapping area is expected to be 500 deg$^2$ and the primary beam of HERA is about 10 degrees, we have assumed that $N_\mathrm{patch} = 5$. If we instead assume a 20\% increase in sky coverage such that $N_\mathrm{patch} = 6$, the S/N ratio increases only by about 10\%. Although this gain is modest and is straightforward to model, it represents a significant increase in the amount of sky area that is jointly covered by the two instruments. As such, it may not be feasible to expand the overlapping area by such a large degree.

\subsection{Deeper Observations}
\label{sec:deeper_surveys}
Alternatively, given additional observing time, it may instead be used to probe the existing sky coverage with greater depth. A fully self-consistent treatment of this choice is subtle, but we briefly discuss some of the high-level effects. By observing the sky for additional time, the limiting magnitude of objects in the survey increases, yielding more observed galaxies. Given that these galaxies are generally hosted in less massive dark matter halos, the associated bias of these objects will decrease (though still remain relatively large, as shown in Figure~\ref{fig:bias_models}), meaning the amplitude of the cross-spectrum will decrease by a small amount. However, given the relatively steep slope of the UVLF for these objects, we expect a significant increase in the number of objects observed.\footnote{Note that a much more significant increase in survey depth than we consider here might have quickly diminishing returns, because many newly-detected faint galaxies are more likely to be in small bubbles, which impede the transmission of Ly$\alpha$ and as a result, our ability to detect the galaxies in Ly$\alpha$.} As discussed above in Section~\ref{sec:s/n}, the cross-spectrum S/N forecasts are limited by shot-noise in the galaxy distribution and the sample variance (i.e., cosmic variance) in the 21\,cm measurements.
Thus by observing more galaxies, we would be able to reduce this contribution to the total noise budget, and increase the significance of detection. 

Figure~\ref{fig:deep_fields} makes this trade-off more concrete. We show the nominal S/N for our 500 deg$^2$ survey as reported in Table~\ref{table:snr} and the expected limiting magnitude of the HLIS, were the detected galaxies have $\mAB~<~26.7$. For our \textsc{ares} models, we are also able to self-consistently model the change in galaxy bias and number density as a function of this magnitude limit (see Figure~\ref{fig:bias_models}). We compute the projected S/N for proposed deeper observations that are 1 magnitude deeper ($\mAB < 27.7$) and 2 magnitudes deeper ($\mAB < 28.7$). To reflect the fact that a smaller sky area is probed, we set $N_\mathrm{patch} = 1$ (i.e., this deeper survey only covers a single patch of 100 deg$^2$). We can see that going 1 magnitude deeper leads to a modest increase in projected S/N, from 12$\sigma$ to 13$\sigma$ for the default set of observing parameters, even at the expense of sky coverage. However, going 2 magnitudes deeper leads to a significant increase in the projected S/N, to 29$\sigma$.

We note also that in principle, these deeper photometric surveys also require deeper grism coverage in the accompanying areas. As shown in Figure~\ref{fig:lya_mab}, going 2 magnitudes deeper in imaging requires roughly 5 times deeper spectroscopy to observe all these sources in Ly$\alpha$. Other studies have considered other possible deep field options \citep[e.g.,][]{Drakos2022}, such as covering 10 deg$^2$ with relatively deep photometry and spectroscopy. In this case, we project a detection of 4.0$\sigma$ for a limiting magnitude of $\mab < 27.7$ (i.e., 1 magnitude deeper than the nominal HLIS), and a 9.0$\sigma$ detection for $\mab < 28.7$.



\section{Conclusion}
\label{sec:conclusion}
In this paper, we explore the prospects for the 21\,cm-galaxy cross-spectrum for upcoming measurements.\footnote{Those interested in understanding more of the implementation details can look at a GitHub repository containing key parts of the calculation: \url{https://github.com/plaplant/21cm_gal_cross_correlation}.}  We show that such a signal is nominally detectable for projected measurements for HERA and the Roman Space Telescope, with a forecasted sensitivity of 12$\sigma$ even under pessimistic assumptions regarding the foreground wedge, i.e., the region of Fourier space unusable owing to foreground contamination. We find that the accuracy of spectroscopically determined redshift values $\sigma_z$ does not dramatically impact the sensitivity of the measurement, though photometrically determined redshifts will not produce a statistically significant detection. In our fiducial forecast, we assume an effective duty cycle of 10\% for the LAE objects detected by the HLS, which has a significant impact on the forecasted sensitivity. If instead the effective duty cycle is 1\%, a significant detection is still possible, though this lower significance can be mitigated with a deeper galaxy survey, less 21\,cm foreground contamination, or an intrinsically larger galaxy bias consistent with the \textsc{BlueTides} predictions. We also explore the impact of varying the reionization history used in our simulations, and find that the effect is not very significant. The one caveat is that for sufficiently late reionization histories (where the midpoint of reionization is $z \lesssim 7$), since the grism on Roman cannot detect $\Lya$ in objects at $z \lesssim 7.2$.

Given the forecast sensitivity for HERA, we expect the 21\,cm measurements to be cosmic-variance dominated for the modes that contribute most to the cross-power spectrum. As such, there is not much additional statistical advantage to be gained from next-generation experiments such as the SKA. However, if the SKA allows for the recovery of signal that is lost to foreground contamination, then the significance of detection may improve dramatically. Indeed, with the projected sensitivity levels, it may be possible to characterize the ionization history, rather than making a simple detection, by dividing the cumulative sensitivity into non-overlapping redshift windows. Additionally, by using additional $k$-bins, it may be possible to constrain the size of ionized bubbles during the EoR. However, we leave a detailed investigation of this possibility for future work.

For the galaxy measurements, we find that the relatively low number of galaxies expected to be measured at high redshift has the largest impact on the overall sensitivity of the measurement. As such, increased observation time of the HLS patch may prove useful for making a more robust measurement. When weighing the potential trade-off between covering more sky area versus performing a deeper survey of the existing footprint, we conclude that a deeper survey is more beneficial. This strategy would lead to a larger number of observed galaxies, thereby decreasing the shot-noise contribution to the overall uncertainty. As this component is the most significant factor, taking steps to decrease it has the largest impact on the overall sensitivity.

We note that our modeling methods in this paper are relatively simple, and do not capture the full interplay between the 21\,cm signal and high-redshift galaxies. In particular, there is no explicit link between the galaxies and the resulting ionization field. Additionally, we have modeled LAEs using a constant factor $f_\mathrm{LAE}$ to account for the fraction of intrinsic LAEs actually observed by Roman. In reality, the observable LAEs will have some spatial and luminosity dependence primarily owing to absorption from neutral regions in the IGM. In the future, it would be beneficial to employ a more self-consistent model. This is of course challenging and will be more computationally expensive than the approach we take here, but will be vital for unbiased inference.

Finally, when looking toward future 21\,cm experiments, it may be useful to consider how projections for cross-correlation measurements such as the ones presented in this paper may provide guidance for array design and construction. When seeking to confirm or independently verify astrophysical and cosmological parameters measured from 21\,cm experiments alone, cross-correlations may provide a key path forward for providing confidence in auto power spectra.

\begin{acknowledgements}
  We thank Steven Furlanetto, Adrian Liu, Andrei Mesinger, and Steven Murray for illuminating discussions regarding this work. We also thank Ronniy Joseph for assistance with the MWA footprints, and Henry Gebhardt for providing the Roman HLS footprint, which was generated by Chris Hirata, projected by Dida Markovic, and is a product of the Roman SIT. This material is based upon work supported by the National Science Foundation under Grant Nos. 1636646 and 2206602, the Gordon and Betty Moore Foundation through grant GBMF5215 to the Massachusetts Institute of Technology, and institutional support from the HERA collaboration partners. HERA is hosted by the South African Radio Astronomy Observatory, which is a facility of the National Research Foundation, an agency of the Department of Science and Innovation. AG's work is supported by the McGill Astrophysics Fellowship funded by the Trottier Chair in Astrophysics as well as the Canada 150 Programme and the Canadian Institute for Advanced Research (CIFAR) Azrieli Global Scholars program. This work used the Extreme Science and Engineering Discovery Environment (XSEDE), which is supported by National Science Foundation grant number ACI-1548562. Specifically, it used the Bridges-2 system, which is supported by NSF award number ACI-1928147, at the Pittsburgh Supercomputing Center (PSC; \citealt{bridges2,xsede2014}). 
\end{acknowledgements}

\bibliography{mybib}
\bibliographystyle{aasjournal}

\appendix
\section{Comparison with \textsc{BlueTides}} \label{sec:appendix}
\begin{figure*}[t]
  \centering
  \includegraphics[width=0.98\textwidth]{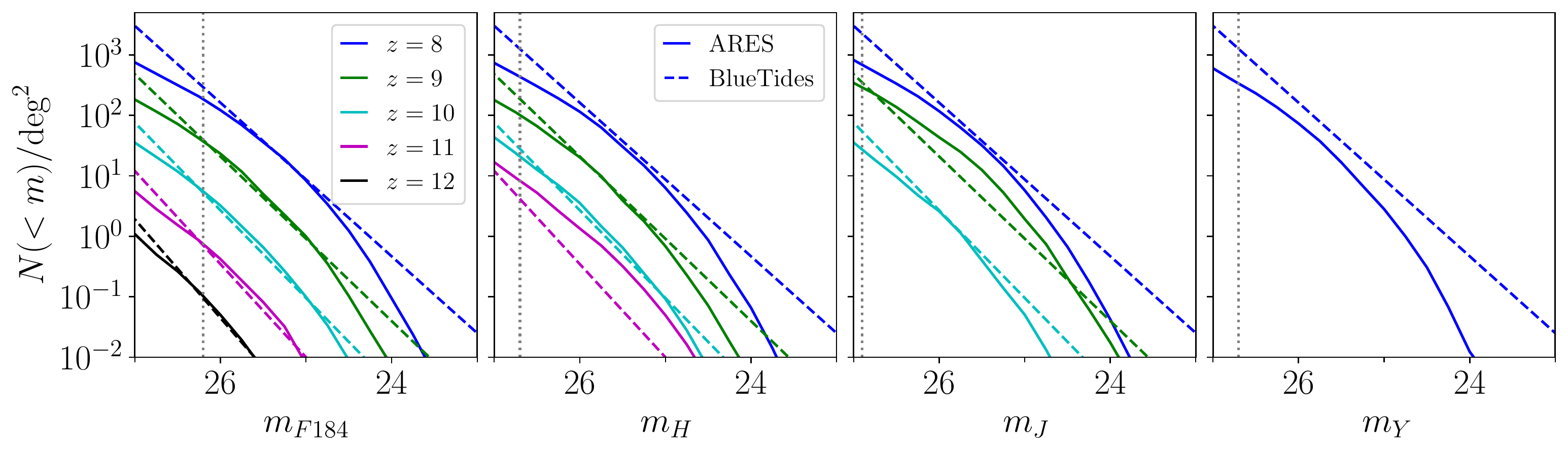}
  \caption{\textbf{Comparison of \textsc{ARES} and \textsc{BlueTides} predictions for galaxy surface densities in the nominal four HLS filters, from reddest to bluest (left to right).} The vertical line in each panel indicates the nominal HLS magnitude limit in that band.}
  \label{fig:ares_v_bluetides}
\end{figure*}

In this section, we show a more detailed comparison between our fiducial model (solid) and the \textsc{BlueTides} (dashed) predictions. First, in Figure~\ref{fig:ares_v_bluetides}, we compare predictions for the cumulative surface density of galaxies detected in each HLS filter. Each panel from left to right indicates a different filter, from the reddest F184 filter to the $Y$ band, which straddles $\lambda \simeq 1 \ \mu\rm{m}$. The magnitude limit is slightly different from band to band, as indicated by the vertical lines. Overall, agreement is fairly good, particularly in the F184 and H bands. The greatest discrepancies are at the factor of $\sim 2-3$ level at $z=8$ for $H$, $J$, and $Y$ bands. This level of disagreement is not unexpected given that, e.g., the UVLFs from \textsc{BlueTides} are not identical to ours (see Figure~\ref{fig:galaxy_properties}).

Though galaxy abundances are fairly similar between models, as shown in Figure~\ref{fig:bias_models}, our predictions for the bias factors are quite different. At $z=8$, \textsc{BlueTides} predicts a linear bias of $b \sim 13$ for HLS galaxies, while our fiducial \textsc{ares} model predicts lower values $b \simeq 8$. This difference is dominated by the different in the stellar-mass-halo-mass relations---at fixed halo mass, \textsc{BlueTides} predicts a lower stellar mass. As a consequence, galaxies of a fixed magnitude will be hosted in higher mass halos, meaning they will be more biased.



\section{Redshift Evolution of the Signal}
\label{sec:appendix_b}

To form a more quantitiative picture of how the signal of interest evolves with redshift, we include an exploration of two useful metrics. First we explore the spherical power spectrum $\Delta^2(k) = k^3 P(k)/2\pi^2$. This quantity is useful to understand because previous studies (e.g., \citealt{lidz_etal2009}) have looked almost exclusively at this quantity rather than the cylindrical power spectra that we consider. We compute the spherical power spectrum $P(k)$ from our cylindrical power spectra at different redshifts. Given our significantly reduced extent along the $k_\parallel$ dimension, there is a relatively restricted range of $k$ values that we are in principle sensitive to. Nevertheless, this quantity is useful for building intuition.

As another measure of the behavior of the signal, we compute the cross-correlation coefficient $r(k_\perp, k_\parallel)$. As with the power spectrum in the main portion of the paper, we compute this quantity using cylindrical coordinates. Quantitatively, $r$ can be expressed as:
\begin{equation}
r(k_\perp, k_\parallel) = \frac{P_{21\times \delta_g}(k_\perp, k_\parallel)}{\sqrt{P_{21,21}(k_\perp, k_\parallel) P_{\delta_g,\delta_g}(k_\perp, k_\parallel)}}.
\label{eqn:r}
\end{equation}
The normalization ensures that $r \in [-1, 1]$. A value of $r=1$ means the two fields are perfectly correlated, and a value of $r=-1$ is perfectly anti-correlated. Given the physical picture of reionization, we expect these quantities to generally be anti-correlated: the galaxy signal comes from regions of high density, and the 21\,cm signal during the bulk of reionization comes from regions of low-density.

\begin{figure*}[t]
  \centering
  \includegraphics[height=0.25\textheight]{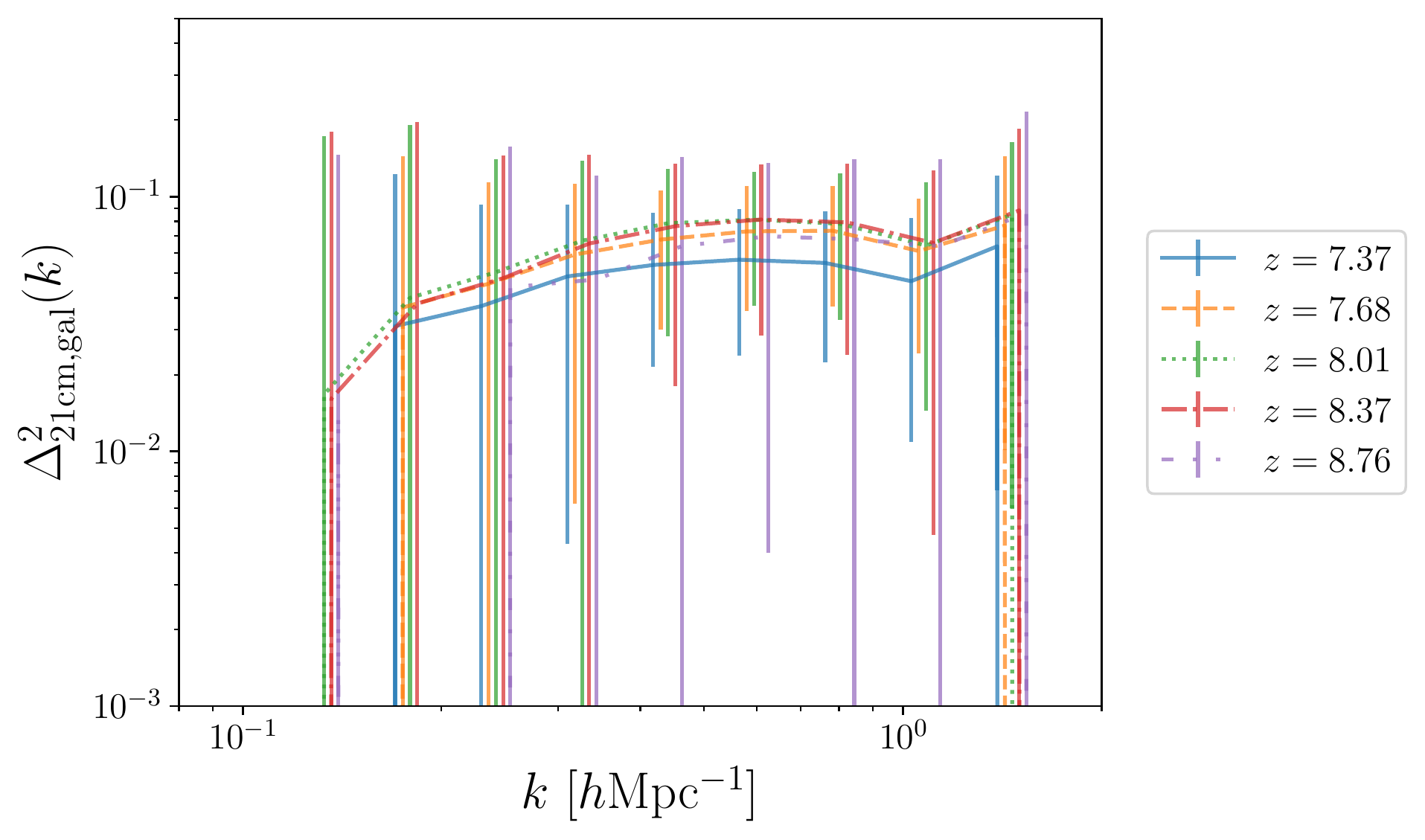}\hspace{10pt}%
  \includegraphics[height=0.25\textheight]{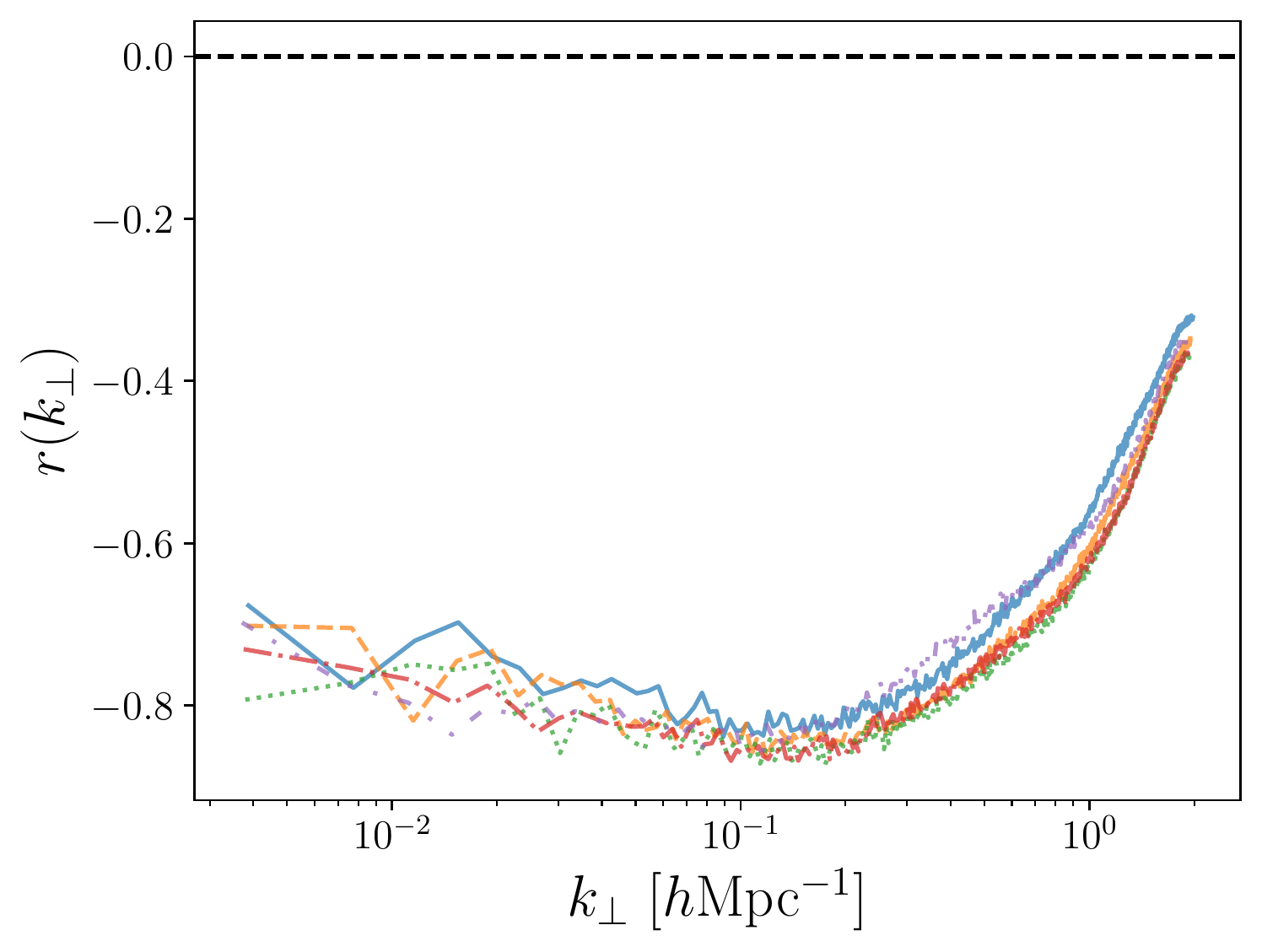}
  \caption{\textbf{Redshift evolution of the cross-power spectrum.} Left: the absolute value of the spherical power spectrum $\Delta^2(k)$ for several different redshift bins. The lines are horizontally offset from each other for visual clarity. We show the 1$\sigma$ error bars for each bin. The corresponding ionization fraction values are $x_i = \{0.63, 0.53, 0.43, 0.33, 0.25\}$. Right: the cross-correlation coefficient $r(k_\perp)$ defined in Equation~(\ref{eqn:r}) for the same redshift values. As can be seen, the fields are highly anti-correlated on all scales, though the anti-correlation decreases on small scales ($k_\perp \gtrsim 1$ $h$Mpc$^{-1}$).}
  \label{fig:sphericalk}
\end{figure*}

Figure~\ref{fig:sphericalk} shows the spherical power spectrum $\Delta^2(k)$ and the cross-correlation coefficient $r(k_\perp)$ for a fixed value of $k_\parallel = 0.1$ $h$Mpc$^{-1}$. We also show the 1$\sigma$ error bars, where we have summed in quadrature the uncertainty described by Equation~(\ref{eqn:xcorr_noise}) for different $(k_\perp, k_\parallel)$ modes that correspond to the same $k$ mode. The cross-correlation signal evolves relatively slowly over this interval. On large scales, the signal peaks near the midpoint of reionization, which is consistent with the behavior of the 21\,cm signal \citep{lidz_etal2008}. Interestingly, the error is smallest at redshifts slightly past the midpoint of reionization. This is a reflection of the fact that the expected number of observed galaxies is larger, meaning the galaxy shot-noise component is smaller. For the cross-correlation coefficient $r(k_\perp, k_\parallel)$, the anti-correlation between the fields is generally quite large on large scale, though not quite perfectly anti-correlated. The decrease toward small scales suggests there is no longer a strong correlation between the galaxy and 21\,cm fields. This result further confirms the desire to work on relatively large scales, such as those probed by HERA and Roman.

\end{document}